%% file: 0-main.tex
\documentclass[sigconf]{acmart}



\graphicspath{{images/}}
\usepackage[a-1b]{pdfx} 
\usepackage{array}
\usepackage{pgfplots,pgfplotstable}  
\usepackage{amsmath}
\usepackage{algorithmic}
\usepackage{enumitem}
\usepackage[ruled,vlined]{algorithm2e}
\usepackage{booktabs}
\usepackage{balance}
\usepackage{multirow}
\usepackage{subcaption}
\pgfplotsset{compat=newest}
\usepackage{balance}
\usepackage{fancybox} % shadow box
\usepackage[most]{tcolorbox}
\usepackage{hyperref}
\usepackage{xcolor}
\newcommand\vldbdoi{XX.XX/XXX.XX}

\newcommand\vldbavailabilityurl{URL_TO_YOUR_ARTIFACTS}
\usepackage{xspace}
 
\newcommand{\sysname}{ENCO\xspace}
\newcommand{\companyname}{Microsoft\xspace}
\newcommand{\icm}{IcM Processor\xspace}

\newcommand{\edit}[1]{\textcolor{black}{#1}}

\newcommand{\md}[1]{}
\newcommand{\yz}[1]{}

\definecolor{teal}{HTML}{00AEB3}

\newcommand\redsout{\bgroup\markoverwith{\textcolor{red}{\rule[0.5ex]{2pt}{0.4pt}}}\ULon}

\definecolor{lightgray}{rgb}{.70,.70,.70}  % define new color
\definecolor{orange}{RGB}{255,127,0}

\newcommand{\tinyskip}{\vspace{1pt}}
\newcommand{\mypar}[1]{\tinyskip\tinyskip\noindent\textbf{#1}\xspace}
\newcommand{\myparr}[1]{\tinyskip\tinyskip\noindent\textbf{#1}\xspace}

\copyrightyear{2026}
\acmYear{2026}
\setcopyright{cc}
\setcctype{by}
\acmConference[KDD '26]{Proceedings of the 32nd ACM SIGKDD Conference on Knowledge Discovery and Data Mining V.1}{August 09--13, 2026}{Jeju Island, Republic of Korea}
\acmBooktitle{Proceedings of the 32nd ACM SIGKDD Conference on Knowledge Discovery and Data Mining V.1 (KDD '26), August 09--13, 2026, Jeju Island, Republic of Korea}
\acmPrice{}
\acmDOI{10.1145/3770854.3783949}
\acmISBN{979-8-4007-2258-5/2026/08}

\begin{document}
\title{\sysname: Deploying Production-Scale Engineering Copilots}

%%
%% The "author" command and its associated commands are used to define the authors and their affiliations.

\author{Yiwen Zhu}
\affiliation{%
  \institution{Microsoft}
  \state{San Francisco}
\country{USA}
}
\email{yiwzh@microsoft.com}

\author{Mathieu Demarne}
\orcid{0000-0002-1825-0097}
\affiliation{%
  \institution{Microsoft}
  \city{Redmond}
\country{USA}
}
\email{mdemarne@microsoft.com}

\author{Kai Deng}
\affiliation{%
  \institution{Microsoft}
  \city{Redmond}
\country{USA}
}
\email{kaideng@microsoft.com}

\author{Wenjing Wang}
\affiliation{%
  \institution{Microsoft}
  \city{Redmond}
\country{USA}
}
\email{wenjwang@microsoft.com}

\author{Nutan Sahoo}
\orcid{0000-0001-5109-3700}
\affiliation{%
  \institution{Microsoft}
  \city{Cambridge}
\country{USA}
}
\email{nutansahoo@microsoft.com}

\author{Hannah Lerner}
\affiliation{%
  \institution{Microsoft}
  \city{Brooklyn}
\country{USA}
}
\email{hannahlerner@microsoft.com}

\author{Anjali Bhavan}
\affiliation{%
  \institution{Microsoft}
  % \department{Azure NW for HC Phynet}
  \city{Burnaby}
\country{Canada}
}
\email{anjalibhavan@microsoft.com}

\author{Divya Vermareddy}
\affiliation{%
  \institution{Microsoft}
  \city{Austin}
\country{USA}
}
\email{dvermareddy@microsoft.com}

\author{Yunlei Lu}
\affiliation{%
  \institution{Microsoft}
  \city{Redmond}
\country{USA}
}
\email{yunleilu@microsoft.com}

\author{Swati Bararia}
\affiliation{%
  \institution{Microsoft}
  % \department{COGS Data - SQL DW}
  \city{Redmond}
\country{USA}
}
\email{barariaswati@microsoft.com}

\author{William Zhang*}
\affiliation{%
  \institution{Carnegie Mellon University}
  \city{Pittsburgh}
\country{USA}
}
\email{wz2@andrew.cmu.edu}
\thanks{*Work done while interning at Microsoft.}

\author{Xia Li}
\affiliation{%
  \institution{Microsoft}
  % \department{COGS Data - SQL DW}
  % \location{5/Mobile}
    \city{Cambridge}
\country{USA}
}
\email{xiali1@microsoft.com}

\author{Katherine Lin}
\affiliation{%
  \institution{Microsoft}
  \city{Redmond}
\country{USA}
}
\email{katlin@microsoft.com}

\author{Miso Cilimdzic}
\affiliation{%
  \institution{Microsoft}
\city{Aliso Viejo}
\country{USA}
}
\email{misoc@microsoft.com}

\author{Subru Krishnan}
\affiliation{%
  \institution{Microsoft}
  \city{Barcelona}
\country{Spain}
}
\email{subru@microsoft.com}

\renewcommand{\authors}{Yiwen Zhu, Mathieu Demarne, Kai Deng, Wenjing Wang, Nutan Sahoo, Divya Vermareddy, Hannah Lerner, Yunlei Lu, Swati Bararia, Anjali Bhavan, William Zhang, Xia Li, Katherine Lin, Miso Cilimdzic, Subru Krishnan}

\renewcommand{\shortauthors}{Yiwen Zhu et al.}

\input{0-0-abstract}

\maketitle

% %%% do not modify the following VLDB block %%
% %%% VLDB block start %%%
% \pagestyle{\vldbpagestyle}
% \begingroup\small\noindent\raggedright\textbf{PVLDB Reference Format:}\\
% \vldbauthors. \vldbtitle. PVLDB, \vldbvolume(\vldbissue): \vldbpages, \vldbyear.\\
% \href{https://doi.org/\vldbdoi}{doi:\vldbdoi}
% \endgroup
% \begingroup
% \renewcommand\thefootnote{}\footnote{\noindent
% This work is licensed under the Creative Commons BY-NC-ND 4.0 International License. Visit \url{https://creativecommons.org/licenses/by-nc-nd/4.0/} to view a copy of this license. For any use beyond those covered by this license, obtain permission by emailing \href{mailto:info@vldb.org}{info@vldb.org}. Copyright is held by the owner/author(s). Publication rights licensed to the VLDB Endowment. \\
% \raggedright Proceedings of the VLDB Endowment, Vol. \vldbvolume, No. \vldbissue\ %
% ISSN 2150-8097. \\
% \href{https://doi.org/\vldbdoi}{doi:\vldbdoi} \\
% }\addtocounter{footnote}{-1}\endgroup
% %%% VLDB block end %%%

% \ifdefempty{\vldbavailabilityurl}{}{
% \vspace{.3cm}
% \begingroup\small\noindent\raggedright\textbf{PVLDB Artifact Availability:}\\
% The source code, data, and/or other artifacts have been made available at \url{\vldbavailabilityurl}.
% \endgroup
\input{0-1-Intro}
\input{0-10-Related}

\input{0-3-Architecture}

\input{0-4-Indexes}

\input{0-5-Backend}
\input{0-5-5-Algorithm}

% \input{0-6-Frontend}
\input{0-7-Evaluation}
% \input{0-8-Deployment}
% \input{0-9-RoadAhead}
\input{0-11-Conclusion}

\bibliographystyle{ACM-Reference-Format}
\balance
\bibliography{sample}

\appendix   
\input{0-12-Appendix}

\end{document}

%% file: 0-0-abstract.tex
\begin{abstract}
 % Incident resolution can be a manual and time-consuming process. For instance, the median time to mitigate a severity 2 incident in Team X at Microsoft is currently 24 hours. 

%  \md{I think we should speak about something else here than that raw data that is across all ICM, and thus relatively meaningless. Saying that Microsoft takes 24 hours for more than half of business-critical incident isn't good.}
% \yz{Yes, edited a bit to broader the focus not only for incident}.

% While on-call duties are inevitable, incident resolution becomes even more daunting due to the obscurity of legacy sources and the pressures of strict time constraints.

Software engineers frequently grapple with the challenge of accessing fragmented documentation and telemetry data, such as Troubleshooting Guides (TSGs), incident reports, code repositories, and internal tools maintained by different teams. 
In this work, we introduced \sysname, a comprehensive framework for developing, deploying, and managing copilots tailored to improve productivity in large scale production scenarios. The framework combines an innovative NL2SearchQuery module with a lightweight hierarchical agentic planner to enable accurate and efficient retrieval-augmented generation (RAG) for code, semi-structured data and documents. These components allow the copilot to retrieve relevant information from diverse sources and invoke the right skills with low latency to answer highly complex technical questions.
% To mitigate documentation gaps that often degrade response quality, \sysname also transforms unstructured incident logs into structured, user-friendly guides.
Since its launch in September 2023, \sysname has demonstrated its effectiveness through widespread adoption, enabling tens of thousands of interactions and engaging over 1,000 monthly active users (MAUs). The system has been continuously optimized based on usage patterns and user feedback, resulting in measurable improvements in response relevance, latency, and user satisfaction.

\end{abstract}

%% file: 0-1-Intro.tex
% !TEX root = 0-main.tex

\section{Introduction}

The daily activities of software developers encompass a broad spectrum of tasks, extending beyond coding, debugging, and testing~\cite{meyer2019today}. Predominantly, engineers allocate the majority of their efforts to ancillary tasks including code review, documentation, and on-call responsibilities~\cite{gonccalves2011collaboration}. These duties necessitate extensive manual searches across disparate resources, ranging from product documentation and troubleshooting guides to telemetry data, system logs, code repositories, and team reports.
% Moreover, engineers invest considerable effort into navigating internal tools that are essential for their daily operations during the arduous task of managing such varied information sources. 
% The challenge is compounded when critical institutional knowledge is isolated or hoarded by senior engineers, resulting in significant knowledge gaps when these key individuals are absent or depart from the organization.
Recent advancements in large language models (LLMs) such as GPT~\cite{gpt}, LLaMA~\cite{llama}, and Gemini~\cite{gemini}, have emerged as powerful tools in this context. 
These models offer significant potential to enhance the engineers’ efficiency by automating mundane tasks, providing swift access to relevant information, and bridging knowledge gaps.
% The deployment of such AI tools not only facilitates more efficient workflow but also ensures centralized access to a diverse array of resources, thereby streamlining the decision-making process for software development teams.
However, despite the availability of advanced AI development platforms such as
OpenAI Assistants API~\cite{openai_assistants_api},
AWS Bedrock~\cite{aws_bedrock},
Google Vertex AI~\cite{google_vertex_ai}, and
Azure AI Agent Service~\cite{azure_ai_agent_service},
which simplifies the creation of engineering copilots by offering access to foundation models, function calling, and API integrations, these solutions remain insufficient for building fully production-grade copilots with proper MLOps support. While these platforms simplify model deployment, they lack key functionalities such as resource management, version control, privacy safeguards, telemetry data collection, CI/CD pipelines, traffic management, and comprehensive evaluation and monitoring mechanisms. 
% These components are essential for ensuring the scalability, reliability, and maintainability of AI copilots in production environments.

Moreover, throughout our experience in developing the solution, we encountered several fundamental challenges:

\mypar{Diverse Data Sources}
To accommodate a broad spectrum of data sources (e.g., telemetry data, historical incident reports, documentation, and source code), a one-size-fits-all retrieval-augmented generation (RAG) solution is insufficient. For instance, historical incident reports often consist of lengthy logs that capture communications between engineers and customers, debugging traces, symptom descriptions, summaries, and external references, resulting in semi-structured data. Technical documentation typically follows a consistent format; however, traditional vector-based retrieval often fails to achieve adequate accuracy due to subtle distinctions within highly technical content. 
% Source code presents yet another challenge, as it is more effectively represented through abstract syntax trees (ASTs), where relational structures play a crucial role. 
 
Although recent advancements in supervised retrieval models (e.g., RAFT~\cite{zhang2024raft}, AR2~\cite{zhang2022adversarialretrieverrankerdensetext}, ARG~\cite{sachan2023questionsneedtraindense},  GraphRAG~\cite{edge2025localglobalgraphrag}, and TREX~\cite{trex}) have significantly improved retrieval performance, they require either the development and fine-tuning of deep neural networks (DNNs) or a significant setup cost.
These approaches are impractical for large-scale self-hosted deployment, as they demand specialized machine learning expertise and substantial infrastructure support.
% for training and deployment, which is not always accessible.

\mypar{Large Number of Skills}
Engineers' daily workflows involve a wide range of complex tasks, requiring a copilot capable of accurate planning and reasoning over an extensive set of skills. However, relying on a single LLM call for skill selection fails to meet the necessary accuracy and consistency standards. While multiple LLM calls can refine decision-making, this significantly increases latency, which is unacceptable for real-time interaction.

Previous research, such as fine-tuning-based approaches~\cite{patil2023gorilla}, has explored training LLMs to generate skill execution plans. However, such methods depend on large quantities of labeled data, making them costly and impractical, especially in dynamic environments where new skills are frequently introduced.

\mypar{Code Indexing}
Code search is a critical capability. Naively treating code as plain text yields poor search accuracy, prompting the development of advanced indexing techniques, many of which leverage compilers. However, modern software codebases are inherently heterogeneous, and they span multiple programming languages, incorporate diverse paradigms, and include non-code artifacts such as configuration files (e.g., YAML, JSON, XML) that define system behavior but fall outside the scope of traditional compilation.
This diversity poses a significant challenge for compiler-based indexers~\cite{treesitter}, which are language-specific and struggle to capture inter-language dependencies and non-source components. The similar challenge was posed for searching Git history.

In this paper, we present \sysname, a universal framework for developing, deploying, and managing copilot applications that support the continuous evolution of a software development team's knowledge base. \sysname is designed to streamline workflows and enhance productivity by focusing on four core areas: 

\begin{itemize}[leftmargin=1em, itemindent=0em]
    \item \textbf{Generalized Development Platform:} Facilitates the rapid creation and deployment of customized copilots by a wide range of contributors.
    \item \textbf{Skills/Agent Integration:} Incorporates an optimized hierarchical agentic framework that enables the integration of new capabilities with high accuracy and low latency.
    \item \textbf{Efficient Retrieval Algorithms:} Enhances retrieval accuracy across diverse data sources with one simple template, including code, semi-structured data, and documents.
    \item \textbf{Streamlined End-to-End Deployment and Lifecycle Management:} Supports a self-hosted model with fully automated deployment pipelines, continuous monitoring, and robust testing tools to ensure maintainability and iterative improvement.
\end{itemize}

\begin{figure}
  \includegraphics[width=\columnwidth, trim=0cm 1cm 0cm 1cm, clip]{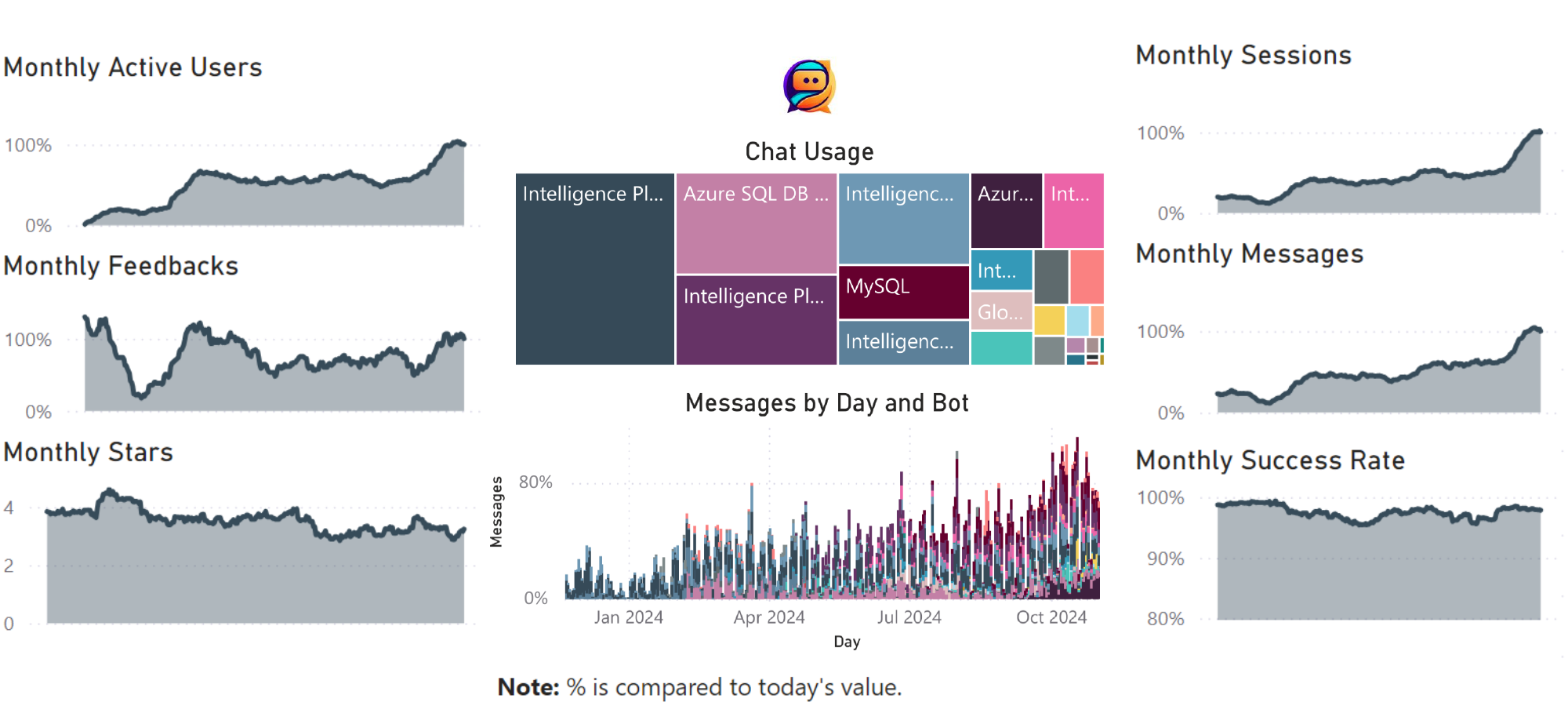}
  \vspace{-0.4cm}
  \caption{Monitoring dashboard for \sysname usage
  % \md{I sanatized this one. Let me know if OK.}
  }
  \vspace{-0.4cm}
  \label{fig:dashboard}
\end{figure}
% Within \companyname, more than 10 teams (referred to as \textit{tenants}) have onboarded to \sysname. 

Using \sysname, dozens of teams across various divisions at \companyname have successfully deployed copilots tailored to their specific roles, adding a comprehensive set of customized skills, making \sysname one of the most popular internal copilot systems. Within the Azure Data organization, we have observed tens of thousands of messages and over 1,000 monthly active users (MAUs) (see Figure~\ref{fig:dashboard}). Across more than 20,000 sessions, conversations average 2.71 rounds, indicating typical multi-turn interactions. 
From several hundred user feedback entries (rated on a star scale), the average rating is 3.6. Interviews with on-call engineers (OCEs) indicate that \sysname significantly reduces the time required for incident triaging \edit{by an average of 10–20 minutes}, demonstrating its practical impact on operational efficiency.

%% file: 0-10-Related.tex
\section{Related Work}\label{sec:related}

% In contrast, our work addresses a broader range of problems, with the copilot supporting general queries beyond incident management to deliver accurate solutions.

% \md{Can I add our very own paper here for RCA? Or should I not? https://dl.acm.org/doi/abs/10.1145/3318464.3386130}\yz{yes, we should~}
% For automating error triage of IcM tickets, AutoTSG~\cite{shetty2022autotsg} constructs executable scripts using TSG documentation, machine learning, and program synthesis to identify relevant troubleshooting steps. Prior works have also focused on automatic debugging and diagnosis of SQL query execution~\cite{zhou2024d}, Spark jobs~\cite{lu2019ladra,lu2017log}, and general root cause analysis for incidents~\cite{roy2024exploring,zhang2024lm,ahmed2023recommending,zhang2024automated,chen2023empowering,mdemarne2020radd}. 
% AskBrain~\cite{chen2020towards} applies language models to automatically categorize and assign incidents to appropriate teams, reducing mitigation time. 
% % However, these efforts primarily target specific categories of issues.
For automating error triage of IcM tickets, AutoTSG~\cite{shetty2022autotsg} generates executable scripts using TSG documentation, machine learning, and program synthesis. Prior work has also addressed automatic debugging and diagnosis of SQL queries~\cite{zhou2024d}, Spark jobs~\cite{lu2019ladra,lu2017log}, and general incident root cause analysis~\cite{roy2024exploring,zhang2024lm,ahmed2023recommending,zhang2024automated,chen2023empowering,mdemarne2020radd}. AskBrain~\cite{chen2020towards} uses LLMs to categorize and assign incidents to appropriate teams, reducing mitigation time.

To improve retrieval-augmented generation (RAG), TILDE~\cite{zhuang2021tilde,zhuang2021fast} and related work~\cite{nogueira2019multi,nogueira2020document} propose document re-ranking using fine-tuned or pretrained models. ReSP~\cite{jiang2024retrieve} iteratively retrieves documents via a ``Reasoner'', though at high latency. Other approaches~\cite{ma2024fine} fine-tune LLaMA for dense retrieval and re-ranking. RAG has also been applied to code completion~\cite{zhang-etal-2023-repocoder,liu2024graphcoderenhancingrepositorylevelcode}, though these systems focus on completing partial code, not answering natural language queries.
% To improve retrieval-augmented generation (RAG) accuracy, TILDE~\cite{zhuang2021tilde,zhuang2021fast} and related works~\cite{nogueira2019multi} propose re-ranking documents using fine-tuned likelihood or embedding models, as well as pretrained relevancy models~\cite{nogueira2020document}. ReSP~\cite{jiang2024retrieve} iteratively retrieves documents using a ``Reasoner'' to assess the current retrieved documents, although this introduces significant latency. \cite{ma2024fine} fine-tunes the LLaMA model as both a dense retriever and a re-ranker in a multi-stage text retrieval pipeline. RAG has also been applied for code completion tasks, such as RepoCoder~\cite{zhang-etal-2023-repocoder} and GraphCoder~\cite{liu2024graphcoderenhancingrepositorylevelcode}. However, these systems typically focus on generating code completions from partially written code, rather than handling natural language queries. 
Recent agentic frameworks support multi-agent coordination and task planning, such as AutoGen~\cite{wu2023autogenenablingnextgenllm} and Letta~\cite{memgpt}, while Gorilla~\cite{patil2023gorilla} targets API selection through LLM-driven function calls. General-purpose platforms like LlamaIndex~\cite{llamaindex}, OpenAI Assistants API~\cite{openai_assistants_api}, AWS Bedrock~\cite{aws_bedrock}, Google Vertex AI~\cite{google_vertex_ai}, and Azure AI Agent Service~\cite{azure_ai_agent_service} offer integration frameworks for LLMs and external tools. Recent advances in graph-based retrieval-augmented generation (e.g., Graph RAG~\cite{edge2025localglobalgraphrag}) leverage structured representations to capture relational information that traditional vector-based methods often overlook.  
However, these solutions lack production-grade deployment capabilities or maintainability. 
% In contrast, \sysname introduces a lightweight, extensible agentic framework designed for real-world engineering environments, supporting multi-turn planning, composable skill execution, and integrated offline preprocessing to optimize both retrieval quality and latency.

% Recent research also explored agentic frameworks to manage complex workflows. Systems such as AutoGen~\cite{wu2023autogenenablingnextgenllm} and Letta~\cite{memgpt} introduce multi-agent coordination patterns for task planning and execution, primarily in research or prototype environments. Gorilla~\cite{patil2023gorilla} focuses on large-scale API selection by prompting LLMs to generate function calls for dynamic skill invocation. LlamaIndex~\cite{llamaindex}, OpenAI Assistants API~\cite{openai_assistants_api}, AWS Bedrock~\cite{aws_bedrock}, Google Vertex AI~\cite{google_vertex_ai}, and Azure AI Agent Service~\cite{azure_ai_agent_service} provides a general-purpose framework for connecting LLMs with external knowledge sources. 
% However, these efforts either (i) lack support for tightly integrated, production-grade deployment pipelines, or (ii) are limited in flexibility. In contrast, ENCO adopts a lightweight yet extensible agentic framework tailored for real-world engineering environments. It supports dynamic multi-turn planning, composable skill execution, and tightly coupled offline preprocessing pipelines, designed to optimize retrieval precision and latency at scale.

%% file: 0-3-Architecture.tex
% !TEX root = 0-main.tex

\section{Architecture}  \label{sec:architecture}

% \begin{figure*}[h]
%   \includegraphics[width=0.8\textwidth]{end2end_new/figures/diagram3.drawio.pdf}  
%   \vspace{-0.3cm}
%   \caption{\sysname architecture}  
%   \label{fig:overviewdiag}  
%   \vspace{-0.3cm}
% \end{figure*}

\sysname comprises 4 primary components: Data Preprocessing, Backend Orchestration, Frontend Services, and Evaluation:  

\textbf{Data Preprocessing [Offline]}: The offline data preprocessing pipelines, created from the same template, are executed at user-defined intervals to draw inputs from various data sources, and generate processed document units (in JSON format) stored in the cloud, served as the fundamental piece to support advanced online retrieval algorithms. 
These pipelines perform additional data enrichment (e.g., keywords or summaries for document chunks). 
Document search indexing services are deployed to index the processed documents, enabling seamless access through API calls from downstream backend (online) modules. Table~\ref{tab:search-schema} illustrates representative fields in the document schema, highlighting which attributes can be searched, sorted, or filtered by the AI Search service~\cite{acs}.

\begin{table}[ht]
\centering
\caption{Portion of the search index schema for Git history.}
\label{tab:search-schema}
\vspace{-0.2cm}
\resizebox{0.9\columnwidth}{!}{%
\begin{tabular}{p{3cm}p{3cm}ccc}
\toprule
\textbf{Field name} & \textbf{Type} & \textbf{Filterable} & \textbf{Searchable} & \textbf{Sortable} \\ \midrule
\texttt{commit\_hash} & String & No & No & No \\ \hline
\texttt{commit\_message} & String & No & Yes & No \\ \hline
\texttt{commit\_date} & DateTimeOffset & Yes & No & Yes \\ \hline
\texttt{pr\_title} & String & No & Yes & No \\ \hline
\texttt{content\_vector} & Vector (3072-d) & No & No & No \\ \bottomrule
\end{tabular}%
}
% \vspace{-0.4cm}
\end{table}

% Table~\ref{tab:index} compares the total indexing time of Azure AI Search against alternative vector databases for SQL documents from six repositories containing approximately 4,000 markdown files, demonstrating Azure AI Search’s high efficiency. Section~\ref{sec:preprocessing} provides a more detailed discussion of this module.

% \begin{table}[h]
% \centering
% \caption{Indexing execution time for SQL from 6 repositories.}\label{tab:index}
% \vspace{-0.3cm}
% \resizebox{0.7\columnwidth}{!}{%
% \begin{tabular}{cc}
% \toprule
% \textbf{Vector DB} & \textbf{Indexing Execution Time} \\ 
% \hline
% Faiss     & $\sim$7 mins  \\ 
% ChromaDB  & $\sim$24 mins \\ 
% Azure AI Search Indexer & $\sim$6 mins \\
% \bottomrule
% \end{tabular}
% }
% % \vspace{-0.8cm}
% \end{table}

\textbf{Backend Orchestration [Online]}: The backend orchestrates the agentic framework using PromptFlow~\cite{pf} where workflows are constructed as a Directed Acyclic Graph (DAG), coordinating input and output exchanges across various modules (e.g., LLM calls, database connections, and other APIs).
% PromptFlow simplifies the integration of LLMs, prompts, and Python code into an executable workflow structured as a Directed Acyclic Graph (DAG), coordinating input and output exchanges across various modules (e.g., LLM calls, database connections, and other APIs).
% It enables straightforward deployment, API and flow monitoring, and greatly reduces maintenance complexity for developers. 
The workflow can be easily deployed in the cloud, serving as a stateless REST API~\cite{onlineendpoint}, optimized for real-time, high-volume, low-latency inference requests.
% , and equipped with extensive logging and debugging capabilities, and can be easily scaled to support thousands of users.
% Additionally, various \textit{context-retrieval skills} can be configured alongside \textit{other skills} to automatically invoke internal toolkits based on the planner’s node output. 
% Sections~\ref{sec:backend} and~\ref{sec:algo} provide further details on these modules.

% PromptFlow offers a collection of development tools that streamline the entire development process for LLM-based AI applications. LLM-based applications essentially involve a sequence of calls to external services such as LLMs, databases, or other APIs, and PromptFlow simplifies the integration of LLMs, prompts, and Python code into an executable workflow. Moreover, it is tightly integrated with the Azure Machine Learning~\cite{aml} infrastructure, facilitating easy deployment, API and flow monitoring, thereby considerably simplifying the maintenance work for developers.  
% We employ a PromptFlow endpoint as our backend engine. 
% , which is integrated with Azure Machine Learning~\cite{aml} and functions as the primary orchestrator. As a component of the Azure OpenAI~\cite{azureopenai} ecosystem, 
% Promptflowm supports easy development, testing, and evaluation of AI-driven prompts, and workflows. 

\textbf{Frontend Services}: The frontend manages user authentication and authorization and handles user interactions. 
It manages session memory, building the full chat history and sending it to the backend's (stateless) REST API to enable a continuous conversation experience. Session state and extracted memory data are stored in the cloud with telemetry data for easy monitoring and review from the API calls. This data is later used for feedback learning~\cite{williampaper} and evaluation (see Section~\ref{sec:eval}).  \edit{We also added an MCP server for backend orchestration, enabling integration with existing MCP clients. More details about the implementation of the components can be found in Appendix~\ref{app:sys}.}

% A web application is deployed with two main user interfaces: a web-based platform and a Microsoft Teams integration, including group chat and session-sharing functionalities.

% Compared to many existing agentic frameworks where session and memory management are handled within the same backend (e.g., AutoGen~\cite{wu2023autogenenablingnextgenllm} or Letta~\cite{memgpt}), implementing a stateless REST API in the backend to operate independently of session state management offers two key advantages: (1) Increased \textbf{development agility} through simplified implementation due to decoupling, and  (2) Seamless backend updates for the copilot, with the flexibility to use \textbf{multiple backend endpoints to manage traffic} for scalability, as session state is handled by the frontend. 

% \begin{itemize}  
%     \item Increased \textbf{development agility} through simplified implementation due to decoupling, and  
%     \item Seamless backend updates for the copilot, with the flexibility to use \textbf{multiple backend endpoints to manage traffic} for scalability, as session state is handled by the frontend.  
% \end{itemize} 
% Section~\ref{sec:frontend} provides further details on this module.

% \textbf{Evaluation}: A comprehensive assessment of the entire system is conducted to monitor and evaluate real user interactions as well as individual modules, ensuring a high-quality chat experience. Section~\ref{sec:eval} discusses this module in more detail.

\textbf{Evaluation}: A comprehensive assessment of the entire system is conducted to monitor and evaluate real user interactions as well as individual modules. 
% The evaluation pipeline supports: (1) Code/prompt change checks to track response quality and prevent regressions; (2) leveraging user feedback to construct golden datasets that guide module development and improve response quality; and (3) using user feedback to enhance documentation retrieval, such as through adaptive re-ranking~\cite{williampaper}. 

% \begin{tcolorbox}[colback=black!5!white,colframe=black!80!white,before upper=\vspace{-.2cm},after upper=\vspace{-.2cm}, left=0.1pt, right=0.1pt, arc=0pt,outer arc=0pt]
%   \myparr{Design Decision:} Using a stateless API for the backend, while delegating memory management to the frontend. 
% \end{tcolorbox}
% \vspace{-0.2cm}

% \md{Note: do we need to introduce PromptFlow earlier here? Or have links to their doc? It's a fairly new tool, so not as ubiquitous.}
% \yz{yes, makes sense. added.}
  
% To deploy the chatbot, a configuration file is required from the tenants to outline several settings. These include: (1) OpenAI settings, for example, the model name and API endpoint, (2) Azure Machine Learning configurations, such as the workspace and subscription names, (3) Azure AI Search settings, such as the service endpoint or the setting for search algorithm,
% % (4) Azure Key Vault configurations, such as the URL for the key vault,
% (4) authentication settings, such Azure identity ID and name, and (5) Promptflow deployment settings, such as the endpoint name, model name, and docker image base. With those configuration files, tenants can easily trigger an Azure pipeline to deploy all the modules. 

%% file: 0-4-Indexes.tex
% !TEX root = 0-main.tex

\section{Data Preprocessing} \label{sec:preprocessing}

In this section, we describe the offline preprocessing pipeline for three data sources: IcM (historical mitigated incidents), documentation, and code repositories. These pipelines build the knowledge base (index) to support advanced context-retrieval skills in the backend, such as \texttt{get\_icm}, \texttt{get\_tsg}, and \texttt{get\_code}. Additional pipelines are created using the same infrastructure to integrate information from other sources, such as Stack Overflow, email or Git history. 

\vspace{-0.1cm}
\subsection{\icm} \label{sec:icm_processing}

\begin{figure}[t]
  \vspace{-0.4cm}
\includegraphics[width=\columnwidth, trim=0cm 5cm 0cm 0cm, clip]{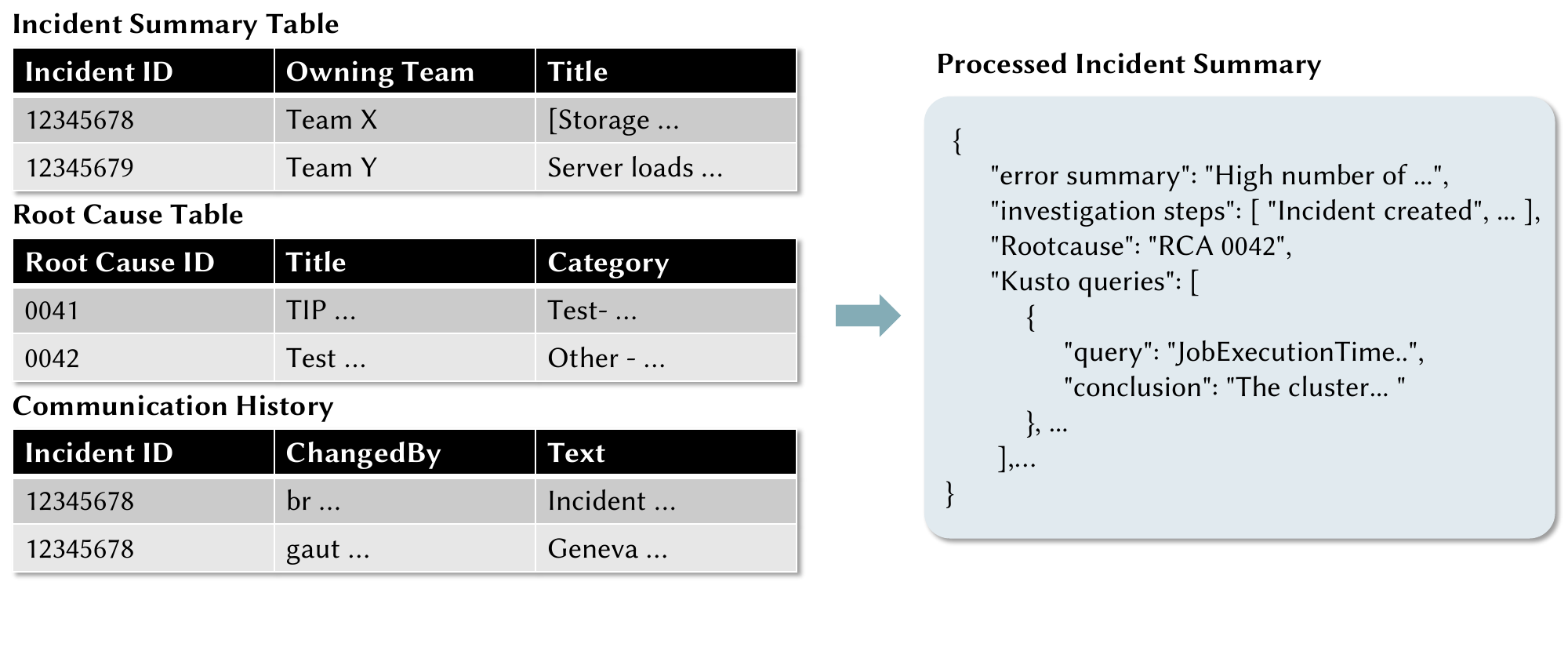}
  \vspace{-0.8cm}
  \caption{Input and output of IcM Processor}
  \label{fig:preprocessing}
  \vspace{-0.4cm}
\end{figure}

The \icm extracts valuable insights from previously mitigated incidents, where oncall engineers may have logged useful discussion, mitigation steps, and commands for troubleshooting. However, as such data can be extremely noisy and lengthy, to better leverage this scattered data, we preprocess the information into a structured format, formulating additional documentation source. The \icm consists of the following customized steps:

\begin{itemize}[leftmargin=1em, itemindent=0em]
    \item \textbf{Data Injection}: The Kusto Connector utilizes secure authentication to access the IcM database. It extracts free-form records based on user-specified team name(s) and date range from multiple data tables (see left of Figure~\ref{fig:preprocessing}). The set of IcMs to be included can be customized through configuration.
    
    % \item \textbf{Data Transformation}: The transformation module employs GPT-4o models to generate structured summaries from free-form text data. This process leverages predefined summarization prompts and curated output templates to extract key fields such as summary, title, mitigation steps, and properties to enable specific lookups (see Section~\ref{sec:icm_improve}). Previous work has demonstrated that LLMs have effectively tackled this task~\cite{pu2023summarizationalmostdead}.

    \item \textbf{Data Transformation}: The transformation module employs GPT-4o models to generate structured summaries from free-form text data. This process leverages predefined summarization prompts and curated output templates to extract key fields such as \texttt{summary}, \texttt{title}, \texttt{mitigation steps}, and \texttt{properties} to enable specific lookups (see right of Figure~\ref{fig:preprocessing}).
    % Previous work has demonstrated that LLMs have effectively tackled this task~\cite{pu2023summarizationalmostdead}.

    \item \textbf{Metadata Enrichment}: Additional metadata is computed, such as a \texttt{helpfulness} score, \edit{generated by LLM,} reflecting the thoroughness of the generated documentation and supporting re-ranking  (see Section~\ref{sec:icm_improve}).
\end{itemize}

This pipeline constructs a structured search index that enables efficient retrieval across multiple fields (e.g., \texttt{summary}, \texttt{mitigation steps}, etc.) to support diverse search strategies during the online phase. Rather than relying on a ``vanilla'' RAG approach, we dynamically construct search strategies based on user input (see Section~\ref{sec:icm_improve}), allowing the system to better align with the intent and context of each user question.

\subsection{TSG Processor}\label{sec:tsg_processing}
% Troubleshooting Guide (TSG) documents provide detailed steps for addressing commonly occurring issues. The processing pipeline described here can be generalized to handle any text-based documentation stored in Git repositories.

The Trouble Shooting Guide (TSG) Processor accesses Git repositories to fetch documentation contents, splits long documents into smaller chunks, and generates embeddings for multiple fields, including document content, titles, and URLs. 
% This enables multi-field hybrid search, improving accuracy and ranking.
Additional metadata, such as document descriptions, referenced IcM IDs, and base64-encoded images are also extracted. The pipeline supports advanced search algorithms, including HyQE question embeddings~\cite{gao2022hyde} and techniques from~\cite{joycepaper}. We conducted a thorough evaluation of documentation retrieval algorithms with different input fields in~\cite{williampaper}.

\subsection{Code Processor}\label{sec:codepreprocessing}
% \yz{? should we include this?}
% \md{I think so! We should add it from the start and put some emphasis somewhere that we allow to bridge from incident to telemetry to documentation to code.}
% To create a search index for code, we employ an internal tool at \companyname, built on the foundation of Tree Sitter~\cite{treesitter}. This tool constructs a comprehensive Abstract Syntax Tree (AST) to extract key components from the code. These components include \texttt{classes}, \texttt{functions}, \texttt{statements}, \texttt{variables}, and their relationships, such as \texttt{used by}, \texttt{uses}, and \texttt{inherits from}. 
% \yz{Hi Mathieu: please let me know if it is OK to include LA-RAG here or if you prefer a standalone paper later on.}

\sysname introduces the \textbf{Language-Agnostic RAG (LA-RAG)}, which leverages LLMs’ multi-language capabilities without relying on compilation.
% Unlike in-memory copilot solutions that process only open files within an Integrated Development Environment (IDE), 
Our offline indexing mechanism enables comprehensive search across large-scale codebases containing thousands of files and millions of lines of code in multiple languages. It achieves this by structuring the codebase as a \textbf{loosely coupled reference graph}, preserving \textbf{one-hop relationships} among code chunks that are most commonly queried by users, and by injecting correlated chunks identified by the LLM in the same search unit.
For the online retrieval, our system is able to accurately determine code construct usage, ensuring context-aware retrieval beyond immediate source files, while avoiding the need of constructing comprehensive knowledge graphs (see Section~\ref{sec:coderetrieval}).

The Code Processor accesses code from Git repositories for data ingestion and tentatively splits long code files into fixed-size smaller chunks. After generating code chunks, we apply the following metadata enrichment and embedding steps:

\begin{itemize}[leftmargin=1em, itemindent=0em]
    \item \textbf{LLM Rechunking}: To prevent functions or classes from being split across chunks, we concatenate each chunk with its five preceding and succeeding chunks, and then employ an LLM to extract complete code segments.
    \item \textbf{Component Relationships}: An LLM is used to identify one-hop related components and attach the corresponding chunks, thereby forming \textit{reference graphs} for broader contextual understanding.
    \item \textbf{Data Augmentation}: The same LLM call also generates a title, a concise description, references used within the code chunk, and summaries of referenced components, thereby enhancing searchability.
    \item \textbf{Embedding Generation}: Embeddings are generated for both code chunks and their associated metadata.
\end{itemize}
% Typically, each copilot indexes approximately \textit{300} to \textit{20,000} code chunks for optimized search and retrieval. 
Code, in this sense, is still chunked using an approximately fixed-size window, while cross-chunk relationships are captured as metadata, enabling relationship analysis during online retrieval (see Section~\ref{sec:coderetrieval}). This approach significantly reduces the number of ``nodes'' compared to the knowledge graph. Furthermore, the online retrieval process can execute multiple search queries sequentially, allowing multi-hop relationship exploration.

%% file: 0-5-Backend.tex
% !TEX root = 0-main.tex
% \begin{figure*}
% \vspace{-0.2cm}
%   \includegraphics[width=0.90\textwidth]{figures/flow.drawio.pdf}
%   \vspace{-0.2cm}
%   \caption{Overview of PromptFlow modules}
%   \label{fig:pf_overview}
% \end{figure*}

\section{Backend Orchestration}\label{sec:backend}

\begin{figure}
\centering
\begin{subfigure}{.5\columnwidth}
  \centering
  \includegraphics[width=\linewidth]{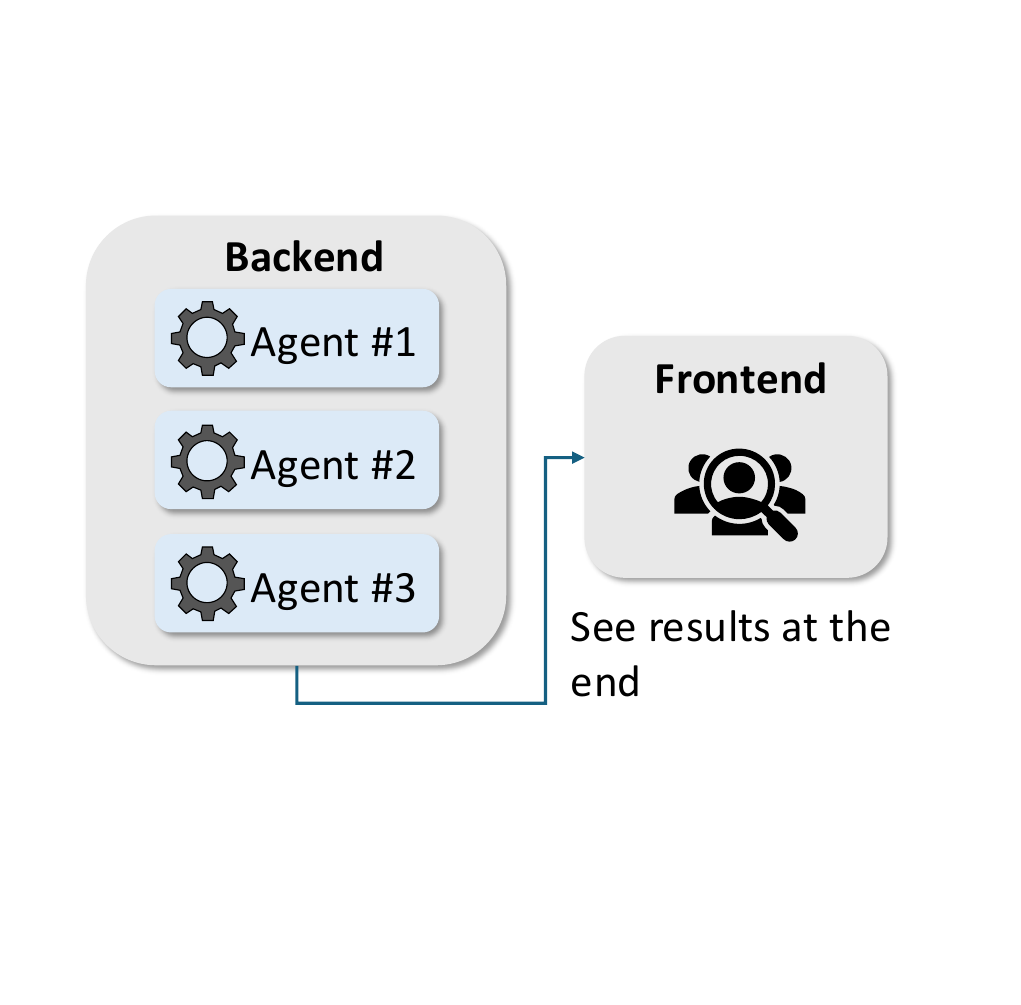}
    \vspace{-0.4cm}
  \caption{A single API call.}
  \label{fig:q1}
  % \vspace{-0.3cm}
\end{subfigure}%
\begin{subfigure}{.5\columnwidth}
  \centering
  \includegraphics[width=\linewidth]{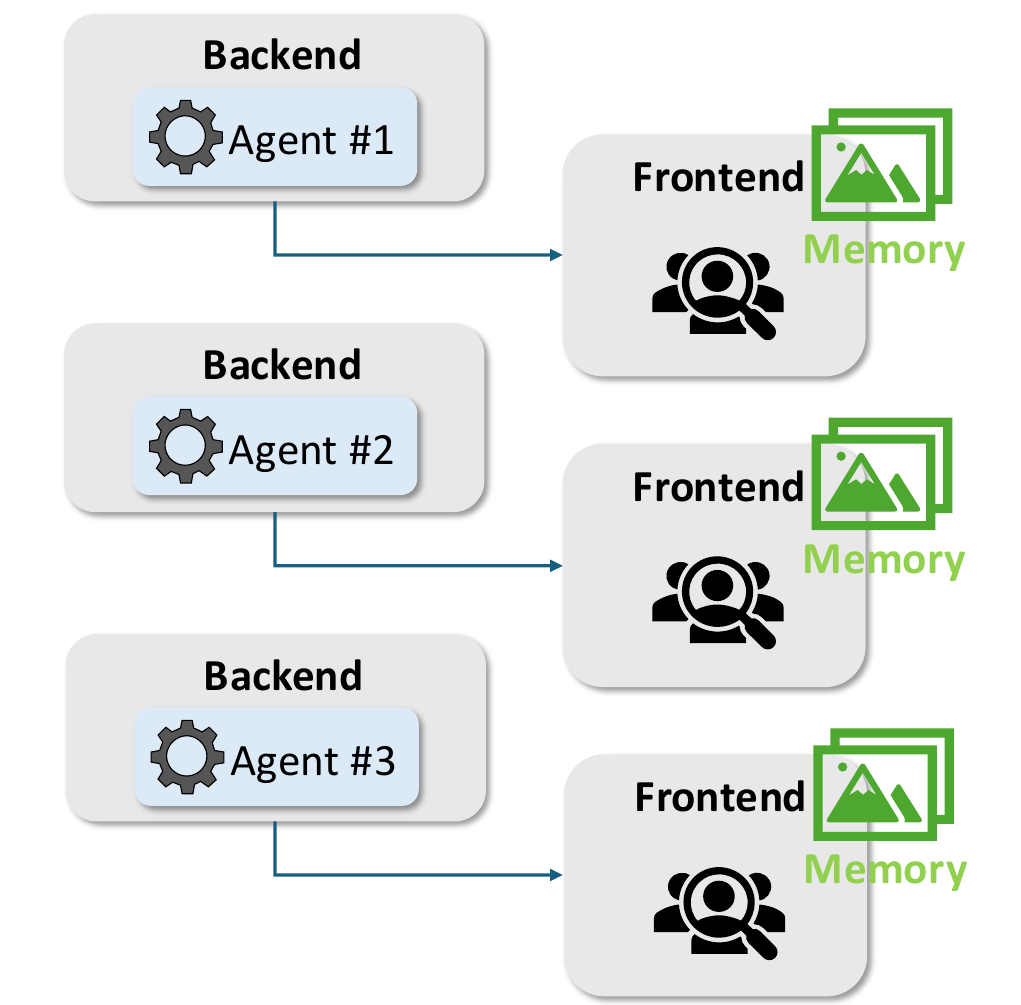}
    \vspace{-0.4cm}
  \caption{Multi-turn interactions.}
  \label{fig:q2}
  % \vspace{-0.3cm}
\end{subfigure}
  \vspace{-0.6cm}
\caption{Frontend-backend interactions}
% \vspace{-0.5cm}
\label{fig:multi-turn}
\end{figure}

Unlike platforms that orchestrate multiple agents within a single frontend API call (left of Figure~\ref{fig:q1}), \sysname executes only one agent per invocation (right of Figure~\ref{fig:q2}). This design eliminates the need for custom mechanisms to handle multi-turn interactions and enables explicit ``reflection’’ after each agent’s execution, offering finer-grained control than free-form chat approaches~\cite{wu2023autogenenablingnextgenllm}. The decoupling of the backend and frontend also increases \textbf{development agility} and maintainability by enabling seamless backend updates for the copilot and supporting scalability through the use of \textbf{multiple backend endpoints to manage traffic}, as session state  (memory) is maintained by the frontend.

\subsection{Agents and Skills}\label{sec:skillagent}

Based on the above considerations, \sysname employs a \textbf{hierarchical agentic planning} approach, dynamically selecting and sequencing agents based on the user question and previous interactions. Instead of treating skills as isolated decision points, the system determines the relevant \textit{agent}, and then, based on the selected agent, determines the appropriate \textit{skills} to invoke sequentially.
% This hierarchical approach reduces complexity, improves efficiency, and maintains flexibility for integrating new skills, ensuring \sysname remains adaptable as capabilities evolve.

% , enabling onboarding teams to seamlessly integrate with various skills or even existing copilots through APIs . Each agent is 

\subsubsection*{\textbf{Agent}} Similar to Claude Code~\cite{claude}, we introduce ``agents'', \edit{where each is a powerful assistant with access to a set of skills, designed to be applied to specific scenarios (see Appendix~\ref{app:agent}).}

% by a YAML file with (1) a description of applicable scenarios; (2) a set of skills the agent can invoke; (3) (optional) an LLM prompt that the agent uses at the end of execution to generate a response, combining all skill outputs as input context; and (4) (optional) a hint for the next agent to be invoked. 
% % Based on the setting of (3), the agent can either invoke an additional LLM call or directly return the skill output as the agent output for this invocation round.

% \begin{figure}[t]
%   % \includegraphics[width=\textwidth]{figures/preprocessing.drawio.pdf}
%   % \vspace{-0.4cm}
%   \includegraphics[width=\columnwidth, trim=0cm 1cm 3.4cm 0cm, clip]{end2end_new/figures/agentic/agent.pdf}
%   \vspace{-0.6cm}
%   \caption{``DiagnosticDocs'' Agent}
%   \label{fig:agent}
%   \vspace{-0.2cm}
% \end{figure}

\subsubsection*{\textbf{Skill}} Each skill extends from a common abstract class and is defined by (1) a detailed description of its functionalities; (2) a specification of its input arguments, such as ``search text'', which need to be extracted from the user input; (3) the execution code (in Python), which may involve API calls, LLM interactions, or any other execution processes; and (4) (optional) skill dependencies, which impacts the execution order, ensuring that required outputs from preceding skills are available as inputs. This information is used to determine the optimal skill execution plan. \edit{Each skill can also optionally output ``hints'' for the next skill/agent to the skill/agent planner, which effectively improve the skill/agent selection for the next round.} 

% For example, the \texttt{get\_tsg} skill requires a ``rephrased question'' as an argument and retrieves relevant documentation via the documentation search API. Its output consists of a list of TSG chunks, which can later be leveraged by the agent’s final LLM call to generate a response. The \texttt{kql.generator} performs NL2KQL functionality and the \texttt{kql.executor} executes it. 
% The skill execution process offers full flexibility, allowing any type of code to be executed.

\subsection{Orchestration}

\begin{figure}[t]
  % \includegraphics[width=\textwidth]{figures/preprocessing.drawio.pdf}
  % \vspace{-0.4cm}
  \includegraphics[width=\columnwidth]{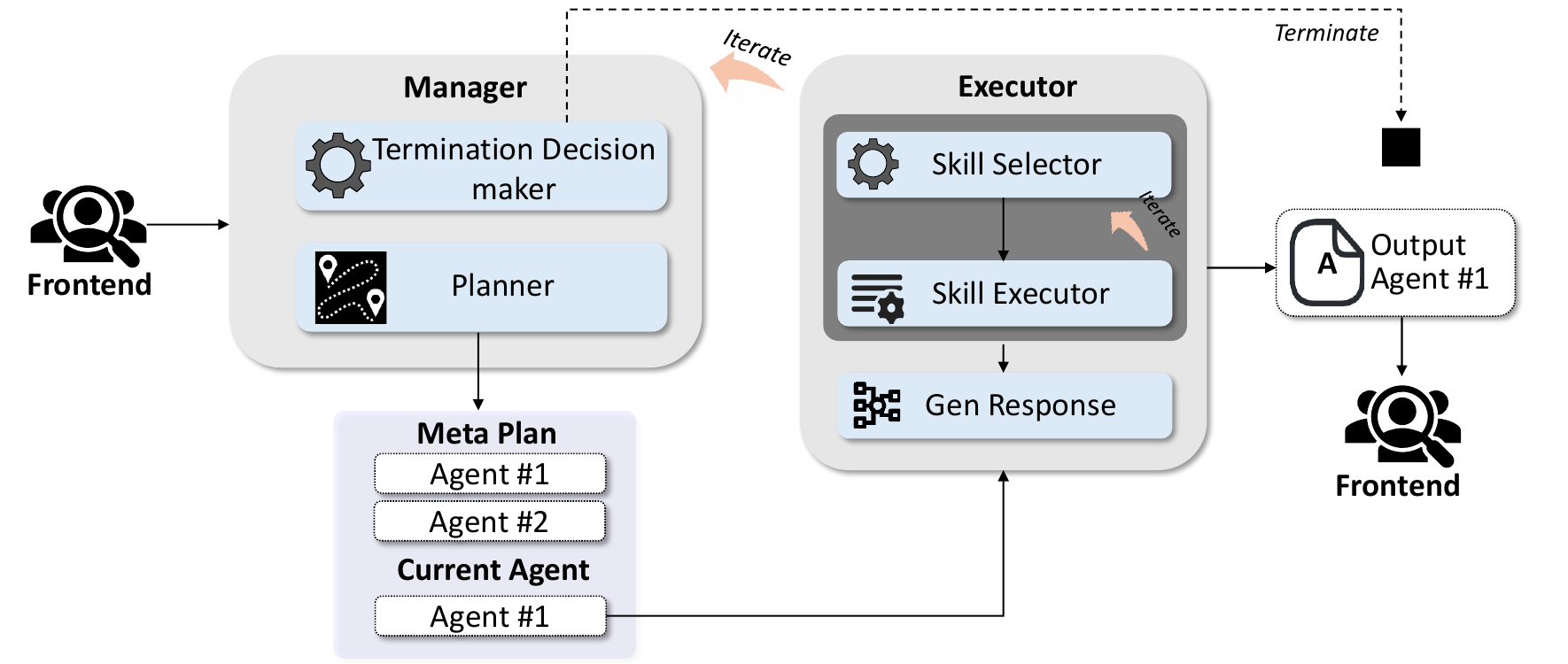}
  \vspace{-0.4cm}
  \caption{Backend orchestration}
  \label{fig:agentic}
  \vspace{-0.2cm}
\end{figure}

Figure~\ref{fig:agentic} illustrates the backend orchestration of the copilot. In each round of backend workflow invocation by the frontend, the \textbf{Manager} module performs two functions in parallel: (1) it determines whether the previous invocation sufficiently answers the user's question (Termination Decision Maker), and (2) it dynamically updates the \textit{meta-plan}, which specifies the sequence of agents to execute and selects the agent to be executed in the current round (Planner).
% the tool-use method is less flexible and requires more human intervention but the free chat method is harder to control and it's more difficult to scope the agents.
The \textbf{Executor} runs the current agent and streams the output to the frontend, triggering the next workflow invocation without user input. This continues until the Termination Decision Maker returns the \texttt{termination} flag. Users can sequentially view individual agent output without waiting for the full meta-plan execution.

% The manager leverages different prompt strategies within its module.

% For each deployment, a \textbf{Config Manager} loads the configuration file to set up the config for each team's agent and skills. For example, the \texttt{get\_tsg} index API for team A differs from team B. All settings are specified in the config files to allow customization.

\subsubsection*{\textbf{Manager}}

% The \textbf{Termination Decision Maker} and the \textbf{Planner} operate in parallel. 
The Termination Decision Maker aggregates the chat history (i.e., the user's questions and the responses from previously executed agents) and determines whether to terminate using an LLM call. 
% The prompt for this call includes detailed termination conditions and examples.
The Planner loads all available agents and compiles their names, descriptions, and hints regarding the next agent into an agent selection prompt. 
% An LLM, invoked via Semantic Kernel~\cite{sk}, generates a sequence of agents to be executed as a meta-plan. 
% During each round of workflow execution, the Planner updates the meta-plan and selects the next agent for invocation.

% The manager, operating within the iterative loop, executes before each invocation of the executor and after the completion of the previous execution. This structure enables support for multiple reasoning paradigms, including the observation and reasoning cycle of ReAct~\cite{yao2022react}, the Chain-of-Thought (CoT) reasoning with self-reflection~\cite{shinn2023reflexion}, and the general \textbf{Plan-and-Execute} approach~\cite{wang2023plan}, which is currently deployed for latency considerations. 
% In our current implementation, most user queries are resolved with fewer than three agents, and the Planner’s decisions are highly accurate.

\subsubsection*{\textbf{Executor}}\label{sec:executor}

% Once the Planner selects an agent for execution, 
The Executor carries out the following steps:
% to ``execute'' the selected agent: 
\textbf{Step 1}: Selects the relevant skill(s) from the agent's available skill set and extracts the corresponding arguments using Semantic Kernel; \textbf{Step 2}: Optimizes the execution plan based on defined skill dependencies, if any, with heuristics to maximize parallelization; \textbf{Step 3}: Executes the skill plan and collects skill outputs; and \textbf{Step 4} (optional): Constructs a prompt based on the agent's definition and makes an LLM call to generate the agent's response. \textbf{Steps 1--3} are executed iteratively until the skill selector decides to terminate, enabling iterative retrieval strategies (for example for code, when multi-hop relationship is desired) and trial-and-error skill execution. 
% \edit{Moreover, in Step~1, we instruct the skill-selection LLM call to generate a ``sub-goal'' for each selected skill, thereby effectively decomposing the complex user query into smaller steps.}

\subsubsection*{\textbf{Memory}}\label{sec:memory}
\begin{figure}[t]
  % \includegraphics[width=\textwidth]{figures/preprocessing.drawio.pdf}
  % \vspace{-0.4cm}
  \includegraphics[width=0.9\columnwidth, trim=0cm 4cm 7.4cm 0cm, clip]{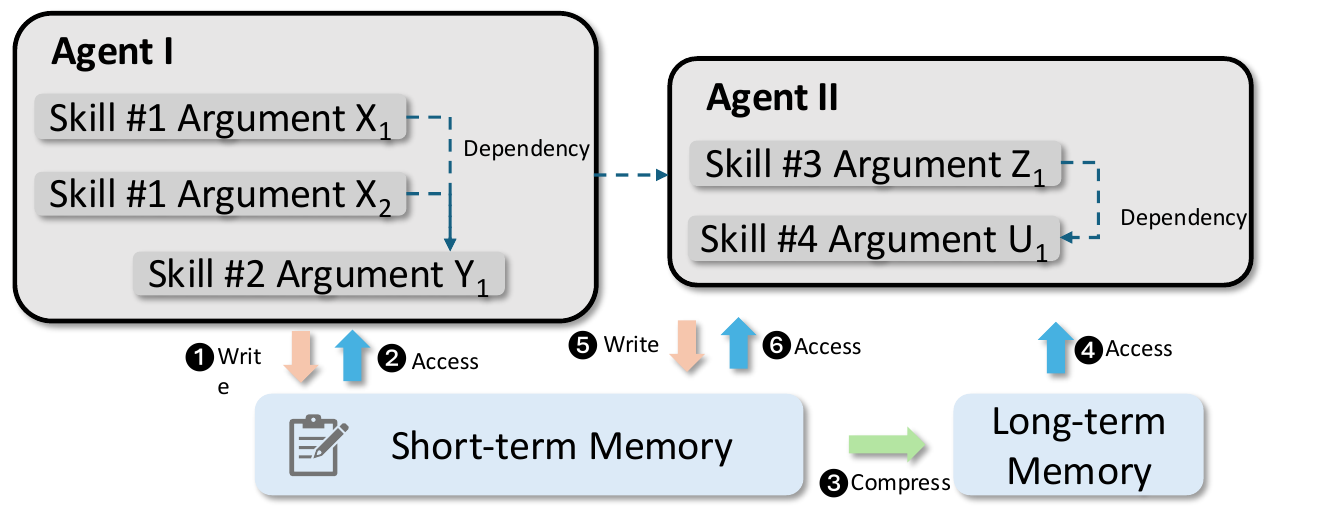}
  \vspace{-0.2cm}
  \caption{Memory management}
  \label{fig:priority}
  \vspace{-0.2cm}
\end{figure}

\begin{figure}[t]
  % \includegraphics[width=\textwidth]{figures/preprocessing.drawio.pdf}
  % \vspace{-0.4cm}
  \includegraphics[width=\columnwidth]{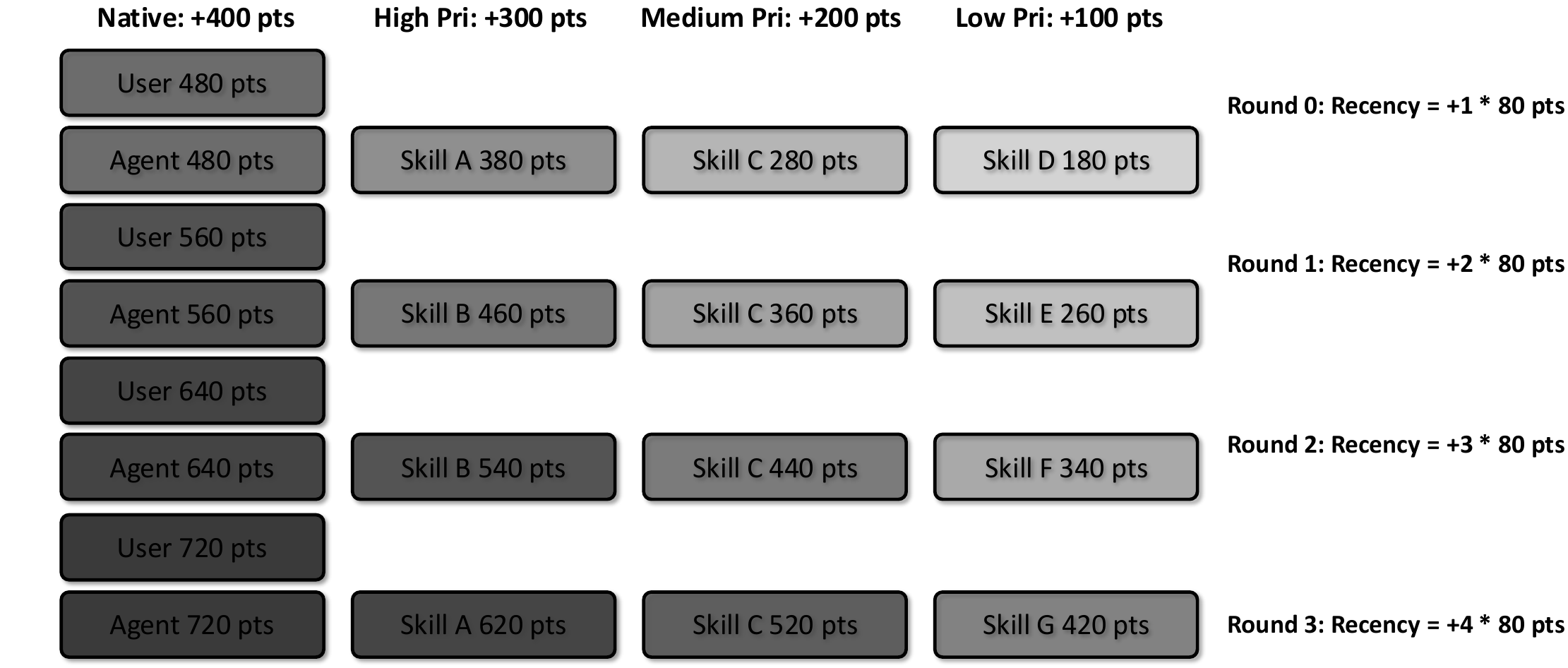}
  \vspace{-0.2cm}
  \caption{Context truncation by priority and recency}
  \label{fig:memory}
  \vspace{-0.2cm}
\end{figure}

\sysname maintains two types of memory: \textit{\textbf{short-term}} and \textit{\textbf{long-term}} (see Figure~\ref{fig:memory}). The short-term memory stores detailed outputs from individual skills, such as the contents of retrieved documents. In contrast, the long-term memory is a compressed abstraction of this raw information and retains only the essential context from the user/copilot conversation.
Within a single agent, each skill writes immediately to short-term memory \edit{(with pre-defined ``priority'' level for the raw skill output)} and can access data produced by previously executed skills. In contrast, subsequent agents access only the long-term memory, constructed at the beginning of the invocation, which preserves the necessary context of prior activities while omitting extraneous details to reduce latency.  \edit{Whenever an LLM call is required, we rank all available context items by combining their
\emph{priority score} and \emph{recency score}. As shown in Figure~\ref{fig:priority}, the
user's latest input and the model's previous response being shown to users are treated as \emph{native} items and
assigned a base score of $+400$ points. Raw outputs from the skills are scored according to their
priority class (\emph{High}: $+300$, \emph{Medium}: $+200$, \emph{Low}: $+100$). In addition,
items from more recent rounds receive an extra recency bonus of $r \times 80$ points, where
$r$ denotes the round index. The resulting combined scores determine which items are retained
when truncating the context given the fix budget (max token length).}

%% file: 0-5-5-Algorithm.tex
\section{Retrieval Algorithms}\label{sec:algo}

In this section, we describe the retrieval algorithms underlying key RAG-related skills, including \texttt{get\_icm}, \texttt{get\_tsg}, and \texttt{get\_code}. 
% Since many chat quality issues originate from incorrectly retrieved data (see Section~\ref{sec:eval_online}), we introduce lightweight and innovative retrieval strategies designed to improve retrieval precision in production environments.

% These enhancements are supported by an accuracy evaluation system, which forms the basis for the continuous evolution of these algorithms, as discussed in Section~\ref{sec:eval}.

\subsection{RAG with NL2SearchQuery}\label{sec:nl2searchquery}

The idea of \textit{NL2SQL}~\cite{nl2sql} is not new. In this paper, we introduce for the first time the idea of \textit{NL2SearchQuery}. Just as SQL queries allow users to retrieve information from structured tables by filtering rows and selecting columns, we propose applying a similar paradigm to searching across diverse documents for Retrieval-Augmented Generation (RAG). A document, much like a row in a table, can be viewed as consisting of multiple ``fields'', such as title, abstract, keywords, body, and references (see right of Figure~\ref{fig:preprocessing}), each of which carries distinct information. By leveraging filtering and sorting based on LLM selected fields and relevance metrics, we can provide search results that align more closely with a user's natural language query and with the most relevant fields to be included as the context.

% In the preprocessing pipeline, documents are enriched with several important metadata fields, creating various opportunities for optimization, such as \texttt{resolve\_date} for IcMs and \texttt{reference} or \texttt{llm\_description} for code. Leveraging the \textsc{NL2SearchQuery} algorithm, we dynamically infer optimal matching criteria or additional filters based on user input to enhance retrieval performance. For instance, user input like ``all the code \textit{referenced by} the class \texttt{SOSScheduler}'' or ``show me \textit{recent} incidents \textit{similar to} Incident XXX \textit{created in} the last two days \textit{involving} login  issues'' can be translated into different search strategies.

% Based on the requirements of the retrieval algorithm, we process different fields for various data sources in the preprocessing pipeline. Similar to a SQL query, we match against specific fields while applying different \textit{filters} or \textit{predicates} to identify the most relevant information. 
% This insight informs the design of our \textbf{NL2SearchQuery} algorithm.
% Based on the needs of retrieval algorithm, we process different fields for different data sources. Similar to a SQL query, we can thus match against different fields and applying various \textit{filters} or \textit{predicates} to pinpoint the most relevant documents. This insight informs the design of our \textbf{NL2SearchQuery} algorithm, 

\begin{figure}
  \includegraphics[width=\columnwidth, trim=0cm 0cm 7.4cm 0cm, clip]{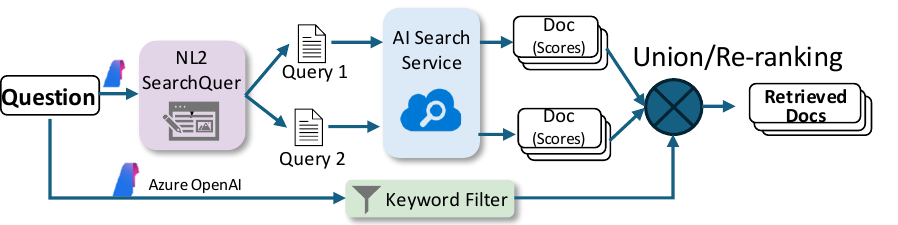}
  \vspace{-0.4cm}
  \caption{RAG with \textsc{NL2SearchQuery}}
  \vspace{-0.4cm}
  \label{fig:nl2searchquery}
\end{figure}

As illustrated in Figure~\ref{fig:nl2searchquery}, we create ``search query'' templates for each retrieval skill and use a large language model (LLM) call to extract key arguments (such as selected search fields, search text, or time ranges) based on user question. Multiple search queries can be executed in parallel using the search service (e.g., Azure AI Search), and the combined result set is re-ranked using various ranking strategies. Additionally, we use the LLM to generate keyword filters from user questions to eliminate irrelevant search results, such as filtering by containment of error codes. In production, we observe that this mechanism significantly improves performance in scenarios where pure vector search is less satisfactory. \edit{This strategy is model-agnostic and can be applied jointly with other embedding models such as ColBERT~\cite{Khattab2020ColBERT} or with traditional methods like TF-IDF/BM25~\cite{chowdhury2010introduction,robertson1995okapi}.}
% The following sections provide a detailed discussion of the retrieval strategies for IcM, TSG, and code, respectively.

\subsection{IcM Retrieval} \label{sec:icm_improve}
% \yz{2. re-ranker for icm and tsg}
% Following the method proposed in \cite{wang2024searching}, we adopt a rewrite-retrieve-re-rank strategy to enhance the precision of our retrieval systems. Our goal is to ensure that the most relevant and useful context is extracted from a diverse set of processed incident reports, as discussed in Section~\ref{sec:preprocessing}.

% Figure~\ref{fig:icm_retrieval} illustrates the IcM retrieval process. 
The \texttt{get\_icm} skill, essential for retrieving historical incident reports, is frequently invoked to answer user questions about troubleshooting procedures or to surface similar past incidents. These questions often mention failure symptoms or specific entities (e.g., server names), which map well to different metadata fields. 
The search query includes the following arguments and filters inferred by LLM:

\begin{itemize}[leftmargin=1em, itemindent=0em]
    \item \textbf{User Intent}: Reformulates the user question by leveraging the full chat context to form precise and self-contained search predicates, used as the matching text for the retrieval.
    
    \item \textbf{Search Field}: Specifies the document field to match against, such as ``\texttt{summary}'', ``\texttt{title}'', ``\texttt{property}'', or ``\texttt{mitigation}''.
    
    % \item \textbf{Search Method}: Chooses the retrieval strategy—hybrid, semantic, simple, or vector-based search~\cite{hybrid}.
    
    \item \textbf{Time Range}: A dictionary with keys like ``\texttt{create\_date}'' or ``\texttt{resolve\_date}'' and values indicating time windows in days.
    
    \item \textbf{Ticket Type}: Specifies the incident type, e.g., ``\texttt{LSI}'' (Live Site Incident), ``\texttt{CRI}'' (Customer Reported Incident), or ``\texttt{ALL}''.
\end{itemize}
% For instance, the query ``Are there any live site incidents created in the last two days involving server testserver1?'' would be translated into: user intent = ``issues on server testserver1'', search field = ``\texttt{property}'', time range = \{``\texttt{create\_date}'': 2\}, and ticket type = ``\texttt{LSI}'' for the search query, using an LLM call.

\myparr{Re-ranking.}
The initial search returns a candidate set of incidents (e.g., 20), which is re-ranked and filtered based on the final keyword filter to select the top $K$ (e.g., 4). Each document $d$ is assigned a \textit{relevance score}: $P(d) = \alpha \cdot \text{IS} + \beta \cdot \text{TS} + \gamma \cdot \text{SS}$, where $\text{IS}$ is the information score, measuring summary quality via token length and the helpfulness score (see Section~\ref{sec:icm_processing}); $\text{TS}$ is the time score, penalizing older incidents; and $\text{SS}$ is the source score, indicating contextual match for team or monitor ID. 
% All scores are normalized and combined to rank the most relevant results.

This retrieval pipeline with fine-tuned configurations based on user feedback, significantly outperforms baseline similarity search using incident titles (see Section~\ref{sec:eval_offline}).

\subsection{TSG Retrieval}
% The \texttt{get\_tsg} function is a critical component of our backend flow. 
In this work, we developed a \textbf{hybrid search algorithm} for the \texttt{get\_tsg} skill, leveraging an advanced preprocessing pipeline detailed in Section~\ref{sec:tsg_processing}, eliminating the need for additional ML training while enhancing both retrieval accuracy and computational efficiency. With the similar NL2SearchQuery algorithm as in Section~\ref{sec:nl2searchquery}, we introduce multiple search queries that run in parallel for this retrieval skill with the following arguments for the search query:

% \mypar{Multi-Field Hybrid Search}
% The hybrid search algorithm utilizes the preprocessing pipeline to extract metadata from each document chunk, including raw content, title, and the associated published URL (as detailed in Section~\ref{sec:preprocessing}). These metadata elements serve as separate embedding targets, and the outputs from all search queries are combined to effectively filter document chunks. We evaluated various search strategies such as vector search, hybrid search, and semantic search, ultimately adopting a hybrid approach that combines embeddings from titles and raw content for optimal performance, based on reciprocal rank fusion (RRF)~\cite{cormack2009reciprocal, microsoft_hybrid_search_ranking}. 

\begin{itemize}[leftmargin=1em, itemindent=0em]
    \item \textbf{Search Text}: The planner rephrases user questions to generate multiple key search arguments.
    % Searches are conducted using both the original user question and this generated search argument. 
    % For queries related to incidents, an additional ``search hint'' is generated based on the incident summary, similar to HyDE~\cite{hyde2023,gao2022hyde}.
    \item \textbf{Search Field}: ``\texttt{title}'', ``\texttt{contents}'' or both.
    \item \textbf{Search Method}: \textit{Multi-Field Hybrid Search} based on reciprocal rank fusion (RRF)~\cite{cormack2009reciprocal, microsoft_hybrid_search_ranking} is used to combine similarity scores from all search fields. 
    \item \textbf{Keyword}: If the user input consists of search keywords (e.g., ``error code \textit{1038}''), additional filter will be added.
\end{itemize}
To further enhance retrieval results, we apply \textbf{Document Filtering} to refine search results. If the highest-ranked document(s) exhibit a significant margin over the following results, only these top documents are retained. We observe that when a small set of highly relevant TSGs is identified, limiting the inclusion of overly detailed or irrelevant documents helps conserve tokens for the LLM calls, reducing response latency (see Section~\ref{sec:eval_offline}).
% Additionally, during preprocessing, teams provide concise \textbf{Repository Descriptions} for each document source. These descriptions are included with the retrieved documents to enhance the LLM’s contextual understanding, improving the quality of the final LLM call.

Future enhancements include integrating more fields in the preprocessing pipelines such as HyDE and Reverse HyDE~\cite{gao2022hyde,hyde2023}, exploring enriched matching fields, incorporating graph structures into the index~\cite{joycepaper}, and utilizing feedback from historical conversations to dynamically refine search rankings~\cite{williampaper}.

% Section~\ref{sec:eval_offline} evaluates the efficiency and effectiveness of these improvements.

\subsection{LA-RAG for Code Retrieval}\label{sec:coderetrieval}

We also apply the NL2SearchQuery algorithm to code search in the \texttt{get\_code} skill for code search given the code has been processed with the preprocessing pipeline (see Section~\ref{sec:codepreprocessing}). 
We design the search query template with following LLM-inferred arguments:

\begin{itemize}[leftmargin=1em, itemindent=0em]
    \item \textbf{Search Text}: Reformulated search elements derived from the user's input, preserving the original intent and contextual meaning.
    \item \textbf{Search Field}: A list defining the search scope for the doc candidates, which may include one or more of the following values:
    \begin{enumerate}
        \item \texttt{title}: Searches the codebase by title, useful for locating specific elements such as method or class definitions.
        \item \texttt{description}: Searches by the LLM-generated description, applicable when users request explanations of the code.
        \item \texttt{content}: Searches by content (the code body), useful for retrieving definitions of specific methods or classes.
        \item \texttt{reference}: Searches by reference, targeting locations where specific methods or classes are used.
        \item \textbf{All}: Searches across all vector indexes of the codebase.
    \end{enumerate}
\end{itemize}
Using few-shot examples, the LLM-generated search query can dynamically determine the most appropriate search criteria/arguments from a rich set of code chunk fields and metadata.
% , effectively handling a wide range of user query types with high accuracy. 
Moreover, based on user requests, LLM can dynamically determine the fields to \texttt{SELECT} from each search unit, including only the fields relevant to the user's query, thereby further reducing the length of the retrieved context.

% Moreover, since skill selection and execution occur iteratively, the same retrieval skill can be invoked in successive rounds, leveraging outputs from previous executions and enabling multi-hop code retrieval.

%% file: 0-7-Evaluation.tex
% !TEX root = 0-main.tex

\section{Evaluation} \label{sec:eval}  
The evaluation system in \sysname is designed to support the continuous improvement of the framework.
% by (1) helping to triage the reasoning behind low-quality responses, (2) evaluating the performance of individual components to guide further development and optimization. 
The evaluation process is divided into two categories: offline and online evaluation.

\subsection{Offline Evaluation}  \label{sec:eval_offline}
% To support ongoing development of the copilot and evaluate changes such as prompt modifications and updates to retrieval algorithms or skill selection mechanisms, we developed an evaluation framework that allows testing in hypothetical scenarios without disrupting the service. 
The offline evaluation focuses on assessing the quality of copilot responses after code changes, acting as a guardrail.

% We observed that LLM evaluation often differs significantly from human judgments, especially those made by domain experts, in terms of relevance and usefulness.  
% \begin{tcolorbox}[colframe=blue!50!black, colback=blue!5!white, width=\columnwidth,before upper=\vspace{-.1cm},after upper=\vspace{-.1cm}, left=0.5pt, right=0.5pt]
% \textbf{Observation:} LLM evaluation often differs significantly from human judgments, especially those made by domain experts, in terms of relevance and usefulness.  
% \end{tcolorbox}  

\begin{figure}[ht]
    \begin{subfigure}{0.48\linewidth}
        \includegraphics[width=\linewidth]{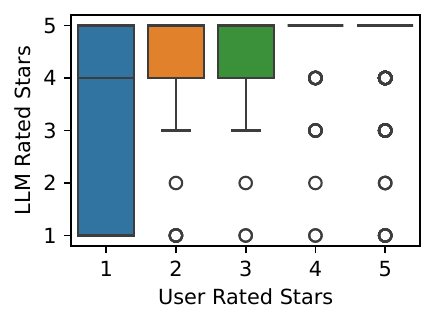}
        \vspace{-2.3em}
        \caption{Boxplot comparison}\label{fig:tsg_eval_cl}
    \end{subfigure}%
    \begin{subfigure}{0.5\linewidth}
        \includegraphics[width=\linewidth]{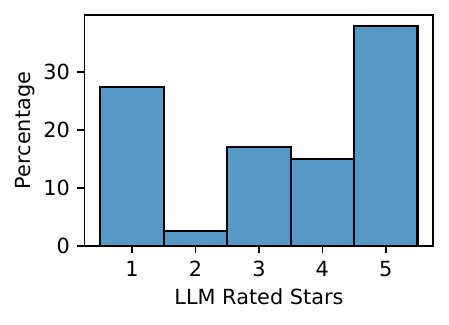}
        \vspace{-2.3em}
        \caption{Among 1-star user ratings}\label{fig:tsg_eval_recall}\label{fig:star2}
    \end{subfigure}%
    \vspace{-0.2cm}
  \caption{Human feedback vs. LLM-predicted ratings}  
  \vspace{-0.4cm}
  \label{fig:star}  
\end{figure}

\myparr{LLM Evaluation.}
We explored the feasibility of employing an LLM to rate copilot responses, simulating human judgment. 
When comparing LLM-generated ratings with human ratings across 660 questions, we found a correlation of only 0.41 (see Figure~\ref{fig:star}), indicating substantial discrepancies. Manual reviews revealed that many responses deemed inaccurate by domain experts were perceived as plausible by non-experts due to subtle errors that only specialists could detect, and responses that received lower ratings from users can be rated higher by the LLM (see Figure~\ref{fig:star2}). This highlights the limitations of using LLMs as evaluators for assessing copilot responses, as discussed in~\cite{NLGEval2023}. As a result, we prioritize using question-answer pairs from user conversations with user ratings as benchmarks for evaluating copilot responses.

\myparr{Evaluation Framework.}
We built a flexible offline evaluation framework for all the evaluation/test suites.
% The framework connects to the telemetry database to download real user queries, copilot outputs, and user feedback, serving as the ``golden answer'' in the evaluation process. 
% The framework features a runner that uses a configuration file to define which evaluators to include in the evaluation pipeline. 
While the end-to-end workflow comprises many interconnected components, our evaluation strategy focuses on \textit{analyzing each sub-component in isolation}. This controlled setup allows us to attribute performance differences specifically to the component under study with low latency. 
Each \textit{evaluator} evaluates one sub-component of the copilot.

% , inheriting from a base class, implements the following functions: 
% data generator, execution function, evaluation function, and report generator.

% (1) \textbf{Data Generator}: prepares input data from telemetry or predefined sources; (2) \textbf{Execution Function}: runs the generated input data through specific components of the chat flow; (3) \textbf{Evaluation Function}: produces evaluation metrics for each processed input and (4) \textbf{Report Generator}: aggregates results from individual samples into a detailed dataset-level report, offering insights into chat flow performance. 

% \begin{figure}  
%   \includegraphics[width=\columnwidth]{end2end_new/figures/eval.drawio.pdf} 
%     \vspace{-0.8cm}
%   \caption{Offline evaluation framework} 
%   \vspace{-0.4cm}
%   \label{fig:eval}  
% \end{figure}  

% Users can extend the framework by adding more evaluators using the provided templates. The evaluation pipeline is integrated into the Azure Pipeline and triggered by Pull Requests (PRs).

\myparr{Evaluators.}
In this paper, we discuss four commonly used evaluators: the Planner Evaluator, the TSG Retrieval Evaluator, the Incident Retrieval Evaluator, and the Similarity Evaluator.

% \begin{figure}  
%   \includegraphics[width=0.6\columnwidth]{end2end/figures/eval/star_boxplot.pdf}  
%     \vspace{-0.4cm}
%   \caption{Comparison of human feedback with LLM-predicted ratings}  
%   \vspace{-0.4cm}
%   \label{fig:star}  
% \end{figure}  

\subsubsection*{Planner Evaluator}
To evaluate the planner's accuracy, we curated 52 questions based on real user questions, each designed to trigger specific skills. For each question, a ``golden answer'' specifies the correct set of skills that should be invoked. We then execute the planner for each question, retrieve the predicted skill list, and compare it with the golden set. 
We report a Boolean metric, \texttt{coverage}, which indicates whether all required skills in the golden set are included among the selected skills.
Planner performance is evaluated across multiple configurations, with five runs per setting. The results indicate that few-shot prompting substantially enhances \texttt{coverage} by approximately 34\% over zero-shot prompting in the flat planner, and by 31 points in the hierarchical agentic planner. Additionally, integrating the hierarchical agentic planner yields a further 4-point improvement when combined with few-shot prompting, reaching up to 99\% coverage and demonstrating exceptional consistency. These gains are also evident in production settings, particularly when dealing with highly customized skills.

To further improve dynamic planning capability, we evaluated the system using 33 questions. Some questions could be resolved with a single-turn skill execution, while others required chaining multiple skills. Using the GPT-4.1 model, the dynamic skill selector achieved a 100\% success rate in selecting the appropriate skill(s). In contrast, the GPT-4o model correctly selected the skill(s) for 29 out of 33 questions, corresponding to an accuracy of 87.9\%, highlighting the benefits of more advanced LLMs.

\subsubsection*{TSG Evaluator}
The TSG Evaluator is designed to assess the accuracy of documentation chunks retrieved by the \texttt{get\_tsg} skill. The first ``golden'' set of TSG documentation is compiled from real user interactions.
% , focusing only on responses rated four or five stars based on the ``reference list" generated at the end of the copilot's response, which includes clickable URLs to the cited documents.
% ; (2) if the complete reference link is missing from the response, we use the output from the \texttt{get\_tsg} module, which provides the full set of retrieved documents, even if not all are cited in the LLM's final answer. 
The second synthetic ``golden'' set is generated by the LLM, which creates synthetic questions for each documentation chunk, allowing us to identify the corresponding TSG documentation that answers those questions, similar to the self-supervised learning~\cite{wang2023solo}. We evaluate the precision and recall by comparing the retrieved documentation set and the golden set. 
% Moreover, we also compute the coverage score to evaluate how well the the retrieved documents ($Q$) ``cover'' the golden set ($G$): $\text{Coverage} = \frac{|G \cap Q|}{|G|}$.

% The evaluation process involves the following steps:

% \begin{itemize}  
% \item Extracting the golden TSG list from conversations with highly rated responses or synthetic data, denoted as $G$.
% \item Running the \texttt{get\_tsg} module for each query to produce a new set of documents, referred to as $Q$. This step yields deterministic results without LLM calls. Optionally, the entire workflow can be executed to capture the final reference list generated by the LLM, denoted as $Q^*$.
% \item Calculating precision and recall metrics by comparing $G$ and $Q^*$:
% \begin{align}
% % \small
% \text{Precision} = \frac{|G \cap Q^*|}{|Q^*|}, \quad
% \text{Recall} = \frac{|G \cap Q^*|}{|G|}.
% \end{align}
% \item Assessing coverage by comparing $G$ with $Q$ to evaluate how well the retrieved documents ($Q$) ``cover'' the golden set ($G$):
% \begin{align}
% \text{Coverage} = \frac{|G \cap Q|}{|G|}.
% \end{align}
% \end{itemize}

\begin{figure}[ht]
  \centering
  \vspace{-0.4cm}
  \includegraphics[width=0.75\columnwidth]{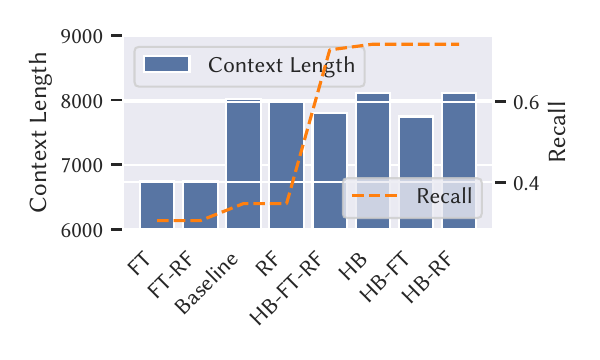}
  \vspace{-0.5cm}
  \caption{Comparison of different retrieval algorithms.}
  \label{fig:tsg_algo}
  \vspace{-0.2cm}
\end{figure}

Figure~\ref{fig:tsg_algo} compares various retrieval algorithms as discussed using SQL copilot's knowledge base with >32,000 documents. The label ``FT'' indicates the use of document filtering, ``RF'' represents introducing search text as a new argument, and ``HB'' denotes hybrid semantic search using multiple fields. We tested 65 questions where the developer team has access to, with 28 sourced from real user data rated above four stars and the remaining 37 generated by LLM based on frequently retrieved document chunks. The results show that hybrid search (``HB-*'') significantly improves recall compared to other methods. Document filtering effectively reduces context length without compromising recall. 
% Although query rephrasing does not notably enhance retrieval accuracy for these specific examples, it proves extremely useful for questions related to Incident Management tickets or those that are poorly formatted by users. For incident-related questions, where the raw query is combined with the incident ticket summary, search results are often suboptimal. However, rephrasing---similar to the HyDE process---enables the generation of optimized search text based on the ticket summary, improving document retrieval. Since generating ticket summaries involves an LLM call, the rephrasing process can be parallelized, avoiding additional latency on the critical path.

\begin{figure}[ht]
    \begin{subfigure}{0.5\linewidth}
        \includegraphics[width=\linewidth]{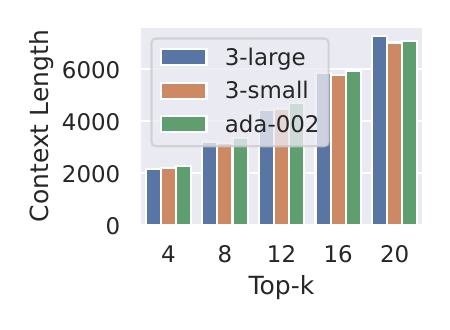}
        \vspace{-2.3em}
        \caption{Context length}\label{fig:tsg_eval_cl}
    \end{subfigure}%
    \begin{subfigure}{0.5\linewidth}
        \includegraphics[width=\linewidth]{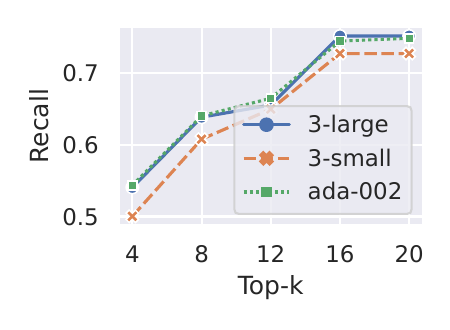}
        \vspace{-2.3em}
        \caption{Average recall}\label{fig:tsg_eval_recall}
    \end{subfigure}%
    \vspace{-0.2cm}
    \caption{TSG evaluator results.}\label{fig:tsg_eval}
    \vspace{-0.4cm}
\end{figure}

We also evaluated retrieval accuracy based on different embeddings. Figure~\ref{fig:tsg_eval} illustrates the comparison of context length and average recall across three most recent models from OpenAI~\cite{embedding}: ``text-embedding-3-large'', ``text-embedding-3-small'', and ``text-embedding-ada-002''. The x-axis represents the configuration settings for the maximum number of documents retrieved. The results show that the ``text-embedding-3-large'' model, with similar context lengths, yields more accurate outcomes. \edit{We also benchmarked \sysname against NUDGE~\cite{zeighami2024nudge}, a state-of-the-art adaptive retrieval algorithm, achieving 18\% improvement in recall~\cite{williampaper}.}
% The same evaluator was also used to fine-tune all the chat flow hyperparameters.

% We conclude that even within the current framework, configuration tuning, including the choice of search algorithm (e.g., hybrid vs. vector search), selection of search queries (e.g., title or content), and embedding models, significantly impacts retrieval performance.

% \begin{tcolorbox}[colframe=blue!50!black, colback=blue!5!white, width=\columnwidth,before upper=\vspace{-.1cm},after upper=\vspace{-.1cm}, left=0.5pt, right=0.5pt]
% \textbf{Observation:} Configuration tuning, including the choice of search algorithm (e.g., hybrid vs. vector search), selection of search fields (e.g., title or content), and embedding models, significantly impacts retrieval performance.
% \end{tcolorbox}

\subsubsection*{Incident (IcM) Retrieval Evaluator}
\begin{figure}[ht]
  \centering
    \vspace{-0.2cm}
  \includegraphics[width=0.78\columnwidth]{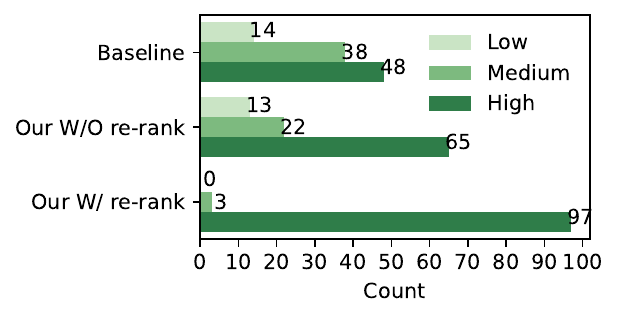}
  \vspace{-0.4cm}
  \caption{Evaluation of similar incident retrieval methods.}
  \label{fig:similar_icm}
  \vspace{-0.2cm}
\end{figure}

We developed a detailed evaluation metric using LLMs by constructing a comprehensive prompt that compares various aspects of incident summaries to assess IcM similarity. For each pair of IcM summaries, we generate a similarity score of ``Low'', ``Medium'', or ``High''. 
When users request similar incidents, we evaluate the similarity between the summary of an incident in the user input and the summaries of retrieved incidents as a proxy for response quality.
% % \md{I changed "a user input incident's" for "the summary of an incident in a user input"}
% 
% The evaluation process involves the following steps:
% \begin{itemize}  
% \item Collecting incident IDs referred to in real user queries.
% \item Extracting the summary of each incident.
% \item Running our \texttt{get\_icm} module to extract corresponding historic similar incident summaries for each input using the algorithm discussed in Section~\ref{sec:icm_improve}.
% \item Assessing the similarity between the input incident and the retrieved incidents.
% \end{itemize}
% 
We executed the evaluator against 100 sampled user questions to select among the >3,000 indexed incidents.
We compare these results against the baseline using vector search based on title similarity. As shown in Figure~\ref{fig:similar_icm}, our retrieval method significantly outperforms the baseline, retrieving more incidents with a ``High'' similarity score. Furthermore, including the customized re-ranking algorithm further improves performance, achieving a ``High'' similarity score for 97 out of 100 incident pairs.

\subsubsection*{Similarity Evaluator}
To maintain consistency in response quality following code modifications, we rerun the copilot to generate new responses and benchmark them against historically high-rated answers, i.e., ``golden-answers''. Specifically, we used LLM to assess the similarity between two answers to the same question, using a 1-5 star rating scale. Each newly generated response aims for a similarity rating of at least 4.0 to meet the release standard.

\subsection{Online Evaluation/Deployment}  \label{sec:eval_online}
% \hl{TODO: add in table showing how the categories are generated: https://eng.ms/docs/cloud-ai-platform/azure-data/azure-data-intelligence-platform/azure-data-dri-copilot/azure-data-dri-copilot/monitoring/telemetry}
% \yz{add maybe some screenshot/a couple sentences from Hannah?}
The online evaluation assesses interactions between the copilot and users in real-world conversations in production. In particular, user ratings of the \texttt{get\_code} skill offer insights into its effectiveness in addressing code-related user questions. Since the introduction of the code search feature, 45 user ratings have been collected, with 71\% awarding 4 or 5 stars, indicating high user satisfactory in practical scenarios.
Representative user inputs include: ``\textit{any SQL Views I can use to query restores}'', ``\textit{from the code, where do we use an identity to get a token to query Arcas / Auto-TSGs}'', and ``\textit{where in the code do we perform On-Behalf-of (OBO) authentication}''. These examples highlight the complexity and specificity of user intent, which goes beyond simple pattern matching approaches as in ~\cite{limpanukorn2025structuralcodesearchusing}.

Table~\ref{tab:search_distribution} summarizes the usage of different code search strategies for 255 user questions. Most questions (61.18\%) use content-based search (\texttt{by\_content}). However, a substantial fraction of questions incorporates additional signals such as title and reference information, either individually or in combination. 
% Notably, 18.04\% of the queries use both title and content features, and 9.80\% employ the comprehensive \texttt{all} strategy. These findings suggest that while content-based search remains the dominant method, diverse strategies leveraging multiple fields (e.g., \texttt{by\_title}, \texttt{by\_reference}) are actively used and contribute meaningfully to retrieval performance. 
This reinforces the importance of supporting flexible and hybrid search strategies in practice to accommodate varied information needs.

\begin{table}[ht]
\centering
\caption{Distribution of code search strategies.}
\label{tab:search_distribution}
\vspace{-0.4cm}
\resizebox{0.8\linewidth}{!}{%
\begin{tabular}{lr}
\toprule
\textbf{Search Strategy} & \textbf{Percentage (\%)} \\
\midrule
\texttt{["by\_content"]} & 61.18 \\
\texttt{["by\_title", "by\_content"]} & 18.04 \\
\texttt{["all"]} & 9.80 \\
\texttt{["by\_content", "by\_reference"]} & 4.71 \\
\texttt{["by\_title", "by\_content", "by\_reference"]} & 1.96 \\
\texttt{["by\_title"]} & 1.96 \\
\texttt{["by\_content", "by\_title"]} & 1.57 \\
\texttt{["by\_reference"]} & 0.78 \\
\bottomrule
\end{tabular}%
}
% \vspace{-0.7cm}
\end{table}

% Conversely, lower-rated interactions tend to involve particularly complex queries that demand advanced reasoning capabilities or cases in which code retrieval is impeded by semantically similar function names. To address such challenges, incorporating feedback-driven learning strategies, as explored in~\cite{williampaper}, may further improve the accuracy and robustness of code search in difficult scenarios.

% \begin{figure}
%   \includegraphics[width=0.64\columnwidth]{end2end_new/figures/eval/star_rating_distribution.pdf}
%   \vspace{-0.4cm}
%   \caption{User ratings for code search skill.}
%   \label{fig:starcode}
%   \vspace{-0.4cm}
% \end{figure}
In addition to user ratings, we employ the following metrics: (1) relevance of the generated answer, scored from 1 to 3, (2) relevance of the retrieved documents, scored from 1 to 3, and (3) groundedness of the answer based on the retrieved documents, scored from 0 to 1. Each metric is evaluated using a specific prompt and an LLM to generate a numeric score. 
An API endpoint has been deployed to asynchronously collect supplementary telemetry data alongside the backend. 
We have categorized the results into five categories to clarify these scores for users (see Table~\ref{tab:online-metrics}).

\begin{table}[h]
\centering
\caption{Answer categories}
\label{tab:online-metrics}
\vspace{-0.4cm}
\resizebox{\columnwidth}{!}{
\begin{tabular}{m{3cm} m{7cm}}
\toprule
\textbf{Category} & \textbf{Details}  \\ \midrule
Relevant-Grounded & Answers with high relevance and grounded in documents. \\ 
Relevant-General & Highly relevant answers based on model's foundational training, not necessarily document-based. \\ 
\begin{tabular}[c]{@{}l@{}}Partially\\ Relevant-Grounded\end{tabular}   & Answers that are partially relevant and grounded in some documents. \\ 
Document Issue & Responses with low relevance due to the absence of appropriate documents. \\ 
Grounding Issue & Responses where relevant documents were found, but the answers were not well-grounded in the documented evidence. \\ \bottomrule
\end{tabular}}
\end{table}

% \md{There was some mentions on stars here, I removed them.}

\begin{figure}
  \includegraphics[width=0.70\columnwidth]{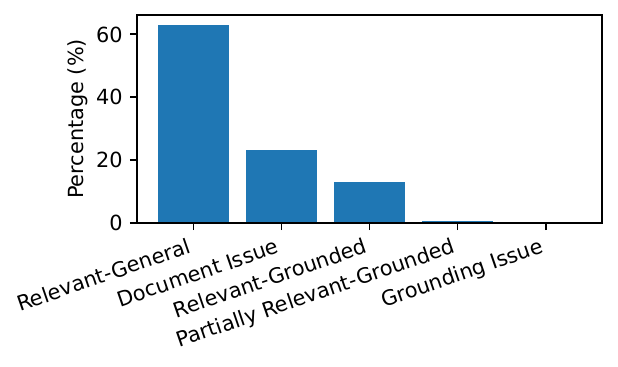}
    \vspace{-0.5cm}
  \caption{Online evaluation category distribution}
  \label{fig:online}
  \vspace{-0.2cm}
\end{figure}

% The online evaluation offers valuable metrics to compare the quality across various deployments for a range of \companyname products. It also enables the measurement of document retrieval quality, which facilitates communication with the owning teams about areas where documentation can be further improved or is currently lacking.
For telemetry data collected through online evaluation, we analyzed a sample of 3,000 messages and categorized the responses as shown in Figure~\ref{fig:online}. The copilot averages 2.2 interaction rounds per session, with 6 rounds at the $95^{\text{th}}$ percentile and 11 rounds at the $99^{\text{th}}$ percentile, indicating that users often engage in \textbf{multi-turn conversations}. \edit{The median response latency is 17 seconds. A detailed breakdown of the cost can be found in Appendix~\ref{app:cost}.}
The most common issue identified was \textbf{Documentation Issues} (23\%), while \textit{hallucination cases (i.e., \textbf{Grounding Issues}) only accounted for 0.1\%}, indicating that hallucination is not the primary concern.
Among conversations flagged as having documentation issues (the 23\%), the root causes can be attributed to three potential factors:
\begin{enumerate}[leftmargin=1em, itemindent=1em]
    \item \textbf{Retrieval skills not being invoked}: The planner fails to trigger the appropriate retrieval skills (46\%).
    \item \textbf{Irrelevant retrieved documents}: The retrieved documents do not align with the user input.
    \item \textbf{Missing documentation}: The relevant document does not exist in the knowledge base.
\end{enumerate}
(2) and (3) account for 32\%, though they are not clearly identifiable.  
Other observed issues include users asking out-of-scope questions (19\%) or providing ambiguous input (2\%). Out-of-scope questions have become a valuable indicator for the team to identify important features and skills to develop.  
Among responses identified as documentation issues, 68\% explicitly state that ``no relevant information was found'', while another 27\% provide a reference list of documents for user verification. This suggests that the LLM effectively detects irrelevant retrieved documents and appropriately disregards them.

Combining user feedback, among the responses where users gave a low rating ($\leq$ 1 star), 44\% were identified with documentation issues. Among these, 93\% explicitly indicated that no relevant information was found. Other low-rating responses where users provided text feedback highlight potential areas for copilot improvement, such as missing functionalities (e.g., inability to execute code) and off-topic responses. These issues may stem from inaccuracies in documentation retrieval or ambiguous user input.

\subsubsection*{Key Learnings}
\textit{1. Fine-tuning is critical.} A substantial effort has been dedicated to optimizing retrieval algorithms for different user scenarios. Even with the same method, different parameter settings can significantly impact performance.
\textit{2. User feedback is invaluable.} Although only 1.7\% of messages include user ratings and 0.5\% contain textual feedback, these inputs have proven to be highly valuable. While the triaging process remains manual, it continues to surface actionable insights, including feature requests, bug reports, and quality issues, that directly inform iterative system improvements (such as the keyword filter). Even with a comprehensive set of testing, we always find ``corner cases" from real user input. Future work involves leveraging the wealth of user feedback to automatically enhance the bot's responses at scale.
\textit{3. First-hand contributors are essential.} Those embedded within domain teams possess deep knowledge of both user needs and domain-specific workflows, enabling them to drive meaningful innovation. The repository includes approximately 200 distinct contributors, with over 90 individuals having made more than 10 commits. This first-hand expertise has been critical to the rapid development and widespread adoption of the framework.
\textit{4. Privacy and security by design.} In a production setting, we place strong emphasis on access control (e.g., who can access particular documents, databases, or skills) and on the responsible handling of sensitive user data. A comprehensive security and privacy review was conducted, covering system architecture, key management, role-based access control, authentication mechanisms (including on-behalf-of token flows), and data handling protocols. Threat models were developed to identify and mitigate risks, and a Responsible AI assessment was performed to ensure compliance and safe deployment. We also designed the monitoring system with privacy safeguards, specifying what information is logged, how long it is retained, and under what conditions it may be reviewed. These measures are critical for maintaining user trust and satisfying enterprise compliance requirements.

% :
% \begin{itemize}
%     \item \textit{The chatbot does not run Kusto scripts.}
%     \item \textit{The command `Get-ServerConfigurationParameters` does not exist.}
%     \item \textit{The main link to EngHub presented is broken.}
%     \item \textit{The real answer is right there: All Trident Sev 0/1/2s (CRI \& LSI) are handled by the Trident SRE team (ICM Service = Power BI, ICM Team = Fabric LiveSite Engineers).}
%     \item \textit{This is not for ARM. It is for the Python-based deployment.}
%     \item \textit{Off-topic answers.}
% \end{itemize}
% , 
% Classification (prototype)
% Relevant-Grounded    15
% Relevant-General     13
% Document Issue        3

% Among the response wehre users give bad rating:

% Classification (prototype)
% Document Issue       16
% Relevant-General     10
% Relevant-Grounded     7

%% file: 0-11-Conclusion.tex
% !TEX root = 0-main.tex
\section{Conclusion}\label{sec:conclusion}

% In this paper, we introduced \sysname, a framework for end-to-end management and deployment of copilots tailored to software engineering tasks. \sysname integrates preprocessing pipelines with tightly coupled front-end and back-end modules to support multi-turn user interactions. 
% % By leveraging historical incident tickets, it bridges documentation gaps with insights from past engineering practices. 
% Its hierarchical, agentic skill planner dynamically selects relevant tools and context. We also developed lightweight retrieval methods based on NL2SearchQuery and Language-Agnostic RAG (LA-RAG) to retrieve relevant incidents, documents, and code snippets.
% A continuous evaluation pipeline monitors retrieval performance and response quality. Deployed across \companyname, \sysname now serves as one of the most popular internal copilot system, improving engineering workflows.
% With increasing confidence in the copilot's responses, future work includes developing an ``autonomous agent'' capable of executing recommended actions without user confirmation to automatically troubleshoot and mitigate incidents and leveraging user feedback to enhance response quality.

% Future work includes enhanced memory management for longer chat histories, and automated mechanisms to learn user preferences and improve documentation retrieval and identify knowledge gaps from conversation data.

In this paper, we introduced \sysname, a framework for end-to-end management and deployment of copilots tailored to software engineering tasks. \sysname integrates preprocessing pipelines with tightly coupled front-end and back-end modules to support multi-turn user interactions. Its agentic skill planner dynamically selects relevant tools and context. We further developed lightweight retrieval algorithm templates based on NL2SearchQuery to retrieve relevant incidents, documents, and code snippets. A continuous evaluation pipeline monitors both retrieval performance and response quality. Deployed across \companyname, \sysname has become one of the most widely adopted internal copilot systems, substantially improving engineering workflows. Looking ahead, future work includes developing an ``autonomous agent'' capable of executing recommended actions without explicit user confirmation to automatically troubleshoot and mitigate incidents, and leveraging user feedback to enhance response quality at scale.

%% file: 0-12-Appendix.tex
% !TEX root = 0-main.tex
\section{Implementation}\label{app:sys}

\begin{figure*}[ht]
  \includegraphics[width=0.9\textwidth]{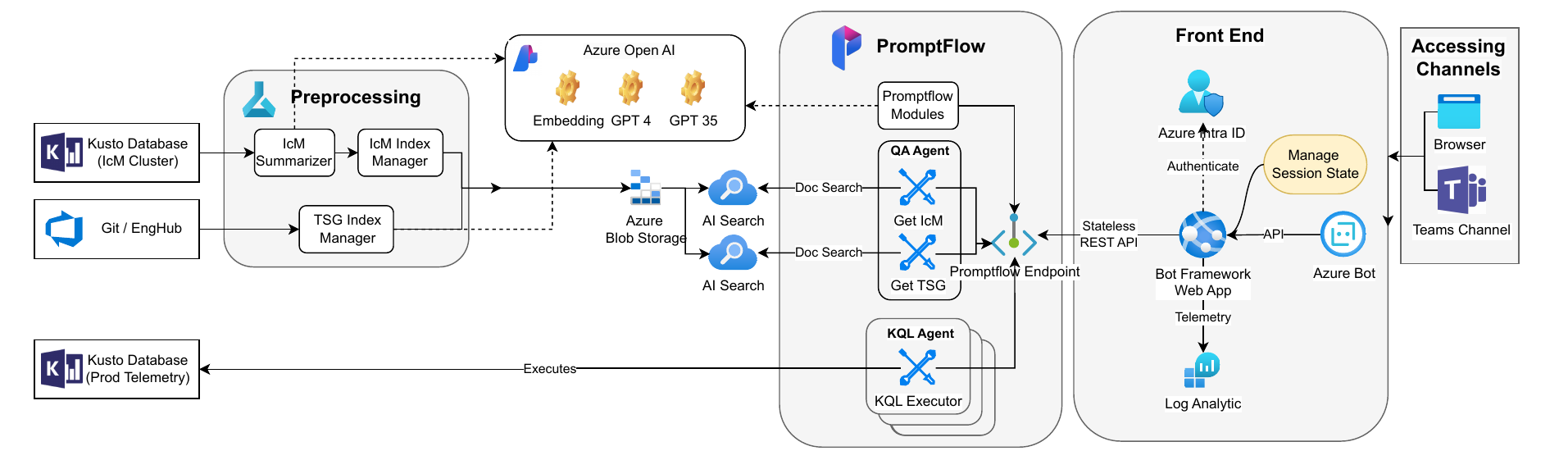}  
  \vspace{-0.3cm}
  \caption{System components of \sysname.}  
  \label{fig:overviewdiag}  
  % \vspace{-0.3cm}
\end{figure*}

Figure~\ref{fig:overviewdiag} highlights the detailed implementation of \sysname and the Azure services powering it.

\textbf{Data Preprocessing}: Data pipelines are orchestrated with \textbf{Azure Machine Learning}~\cite{aml}, pulling from Git, Wikis, and Kusto databases to produce summarized documents stored in Azure Blob Storage. Multiple Azure AI Search indexes~\cite{acs} reference these blobs for retrieval by downstream modules.

\textbf{Backend Services}: The Planner–Skill workflow is managed with PromptFlow~\cite{pf}, integrated into Azure Machine Learning for deployment as stateless REST endpoints~\cite{onlineendpoint}. This supports low-latency inference with monitoring, logging, and debugging.

\textbf{Frontend Services}: User sessions and authentication are handled via Azure Active Directory. Session memory is stored in Azure Blob Storage, while telemetry is captured by Azure Log Analytics. A web app connected to the Azure Bot Framework provides interfaces via browser and Microsoft Teams.

\section{Agent Definition Example}\label{app:agent}

Each agent is defined by a YAML file with (1) a description of applicable scenarios; (2) a set of skills the agent can invoke; (3) (optional) an LLM prompt that the agent uses at the end of execution to generate a response, combining all skill outputs as input context; and (4) (optional) a hint for the next agent to be invoked. 
% Based on the setting of (3), the agent can either invoke an additional LLM call or directly return the skill output as the agent output for this invocation round.

\begin{figure}[ht]
  % \includegraphics[width=\textwidth]{figures/preprocessing.drawio.pdf}
  % \vspace{-0.4cm}
  \includegraphics[width=\columnwidth, trim=0cm 1cm 3.4cm 0cm, clip]{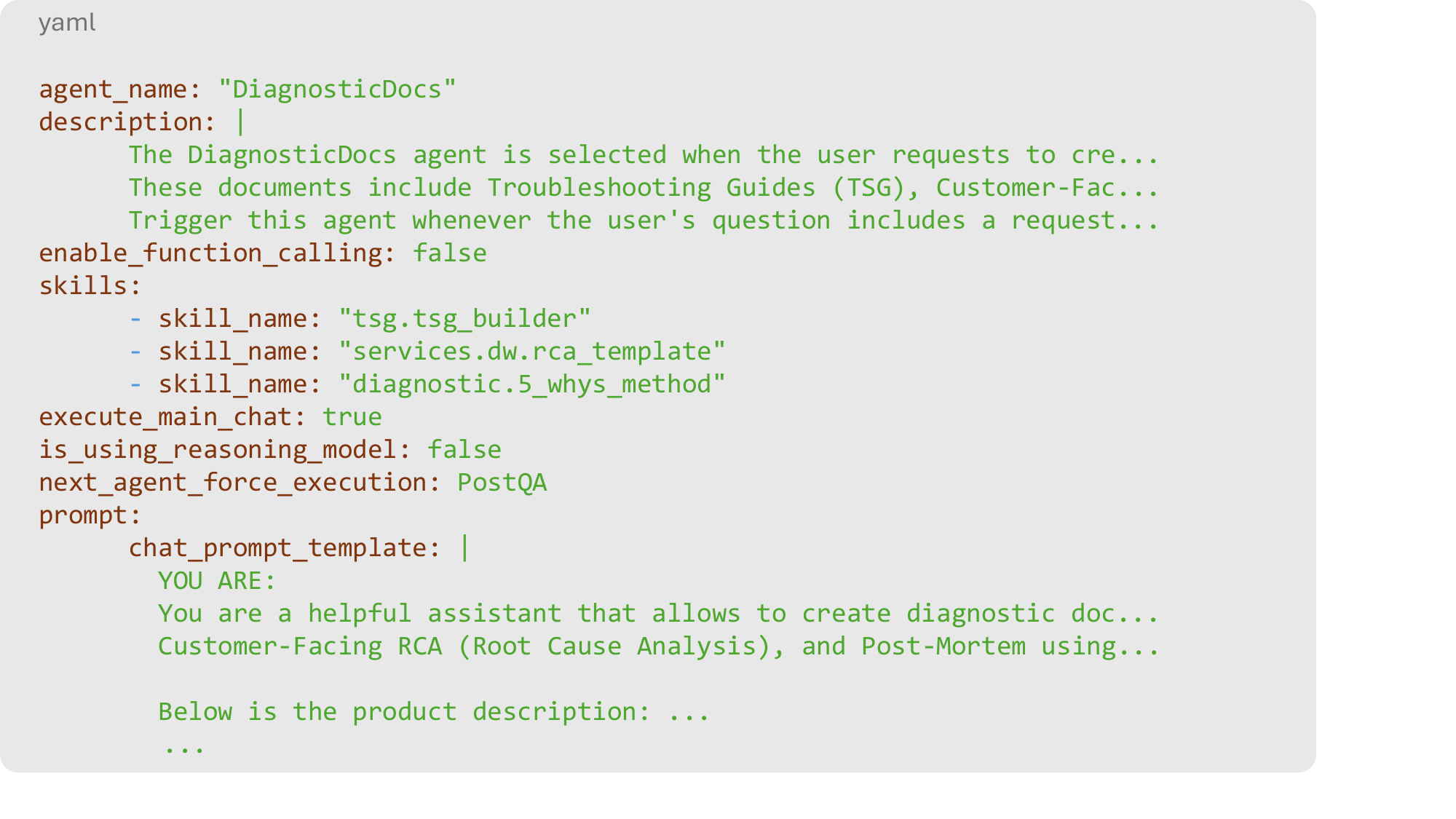}
  \vspace{-0.6cm}
  \caption{``DiagnosticDocs'' Agent}
  \label{fig:agent}
  \vspace{-0.2cm}
\end{figure}

\section{Cost Breakdown}\label{app:cost}

\noindent The primary cost categories for a typical deployment of \sysname are:
\begin{itemize}
  \item Cognitive Services (Azure OpenAI)
  \item Azure VMs (for processing data and hosting models)
  \item Azure Storage (to store prepared data)
  \item Azure AI Search (for indexing and document retrieval)
  \item Azure Application Services (for hosting the frontend)
\end{itemize}

Table~\ref{tab:cost-breakdown} shows the detailed breakdown considering a deployment serving \textasciitilde120 users and \textasciitilde2{,}000 interactions per month as a back-of-the-envelope example.

\begin{table}[h]
  \centering
  \caption{Monthly Cost Breakdown (USD)}
  \label{tab:cost-breakdown}
  \begin{tabular}{l r}
    \toprule
    \textbf{Cost Component} & \textbf{Monthly Cost (USD)} \\
    \midrule
    VMs (on AML)          & \$259 \\
    Cognitive Services    & \$288 \\
    Storage               & \$77  \\
    Cognitive Search      & \$50  \\
    Load Balancer         & \$44  \\
    Container Registry    & \$30  \\
    Azure App Services    & \$27  \\
    \midrule
    \textbf{Total}        & \textbf{\$775} \\
    \bottomrule
  \end{tabular}
\end{table}

% \begin{table}[h]
%   \centering
%   \caption{Unit Economics}
%   \label{tab:unit-economics}
%   \begin{tabular}{l l l}
%     \toprule
%     \textbf{Metric} & \textbf{Value} & \textbf{Cost per Unit (USD)} \\
%     \midrule
%     Users     & 120   & --                     \\
%     Sessions  & 800   & \$0.95 per session     \\
%     Messages  & 2{,}000 & \$0.36 per message   \\
%     \bottomrule
%   \end{tabular}
% \end{table}

Costs scale non-linearly with usage; higher utilization typically lowers the cost per message. Additionally, certain VM resources can be scaled down to save cost depending on usage patterns. The total monthly cost in this example is approximately \$775.